\documentclass{aa}
\usepackage{graphics}
\begin{document}
\title{The new sample of giant radio sources\\
II. Update of optical counterparts, further spectroscopy of identified faint
host galaxies, high-frequency radio maps, and polarisation properties of
the sources}  
\author{J. Machalski, M. Jamrozy, S. Zola, and D. Koziel}
\offprints{J. Machalski}
\institute{Astronomical Observatory, Jagellonian University,
ul. Orla 171, PL-30244 Cracow, Poland\\
email: machalsk@oa.uj.edu.pl}
\date{Received ....; accepted ...}
\abstract
{Our sample of giant radio-source candidates, published in Paper I,
is updated and supplemented with further radio and optical data. In this paper we
present: (i) newly detected host galaxies, their photometric magnitude, and
redshift estimate for the sample sources not identified yet, (ii) optical spectra
and spectroscopic redshift for the host galaxies fainter than about $R\approx 18.5$
mag taken with the Apache Point Observatory 3.5m telescope, and (iii)  the VLA
4.9 GHz total-intensity and polarised-intensity radio maps of the sample
members. In a few cases they reveal extremely faint radio cores undetected before,
which confirm the previously uncertain optical identifications. The radio maps are
analysed and the polarisation properties of the sample sources summarised.
A comparison of our updated sample with three samples published by other authors 
implies that all these four samples probe the same part of the population of
extragalactic radio sources.
There is no significant difference between the distributions of intrinsic size
and radio power among these samples. The median redshift of 0.38$\pm$0.07 in our
sample is the highest among the corresponding values in the four samples, indicating
that the angular size and flux-density limits in our sample, lower than those for
the other three samples, result in effective detections of more distant, giant-size
galaxies compared to those detected in the other samples.
This sample and a comparison sample of `normal'-size radio galaxies will be used
in Paper III (this volume) to investigate of a number of trends and correlations
in the entire data.  
}
\authorrunning{J. Machalski et al.}
\titlerunning{New sample of giant radio sources}
\maketitle

\section{Introduction}

Existing analytical models of the dynamical evolution of classical double radio
sources (e.g. Kaiser et al. 1997; Blundell et al. 1999; Manolakou \& Kirk 2002)
predict that the largest sources must represent a very late phase of their
lifetimes. In fact,
the most extended structures are observed at rather low redshifts, confirming the
effect of `youth-redshift degeneracy' postulated by Blundell et al. Therefore,
a search for `giant' radio sources at high redshifts can reveal environmental
conditions in denser, not yet fully virialised, extragalactic medium very likely
governing the size evolution. On the other hand, their highest energy content, the
large ordered magnetic field structures, the typical absence of strong large-scale
shocks, and very low upper limits on their internal thermal plasma densities,
make them - according to Kronberg et al. (2004) - `prime and unique laboratories
for constraining the plasma processes that accelerate relativistic electrons
within large extragalactic volumes'.

For proper statistical studies of the above conditions and constraints, numerous
complete samples of extended radio sources are necessary. The above idea was followed
at the same time by independent teams (Schoenmakers et al. 2001; Lara et al. 2001;
Machalski, Jamrozy \& Zola 2001 [hereafter Paper I]). All of these three samples
were selected from the large radio surveys: WENSS (Rengelink et al. 1997) and NVSS
(Condon et al. 1998). Recently another sample of giant-size radio sources, selected
from the SUMSS (Bock et al. 1999) survey in the southern sky hemisphere, has been
published by Saripalli et al. (2005).

In Paper I we presented a sample of 36 large angular-size, double radio sources,
in principle with the FRII-type (Fanaroff \& Riley 1974) morphology, selected from
the VLA 1.4-GHz sky surveys: NVSS and the first part of FIRST (Becker et al. 1995).
These two surveys provide radio maps of the sky with two different angular resolutions
(45$\arcsec$ and 5$\arcsec$, respectively) at the same observing frequency, allowing
for (i) effective removal of confusing sources, (ii) reliable recognition of
double-source identity and determination of its morphological type, and (iii) (for
many of the sample sources) determination of the radio core component necessary for
the proper identification of the selected source with its host optical object
(mostly a galaxy).

In order not to miss sources larger than 1 Mpc\footnote{A conventional
lower limit for the projected linear extent of giant radio sources
if $H_{0}$=50 km\,s$^{-1}$Mpc$^{-1}$ and $q_{0}$=0.5 used in most of
previously published papers} at a redshift above 0.3, the sample candidates
were selected according to the following selection criteria: each source
lies within the sky area of 0.47 sr, limited by
07$^{\rm h}$20$^{\rm m}$$<$RA(2000)$<$17$^{\rm h}$30$^{\rm m}$ and
+28$\degr$5$<$Dec(2000)$<$+41$\degr$0, and also has the FRII (or eventually FRI/II)
morphological type and an angular size larger than 3$\arcmin$, as well as the 1.4-GHz
flux density 30 mJy$<$$S_{1.4}$$<$500 mJy. The last criterion sensibly excluded
the already well-known  giants lying in the sample area (e.g. B\,0854+399,
3C\,236, B\,1358+305).

In Paper I we presented (i) high-resolution VLA observations at 4.86 and 8.46 GHz,
which allowed us to detect very weak radio cores in a number of sample sources
undetected during the FIRST survey and to identify them with their optical
counterparts, (ii) deep optical detections of 9 faint host galaxies beyond the
magnitude limits of the existing data bases (Palomar Observatory Sky Survey [POSS],
Digitized Sky Survey [DSS]) and their crude photometry, and (iii) low-resolution
optical spectra that provided redshifts for 11 identified host galaxies brighter
than about $R\approx 18.5$ mag (for the other 5 galaxies their redshifts were known
prior to our observations).
Consequently, there was no radio core detection for 8 sample sources, uncertain
optical identification for 6 sources, no identification for 4 sources with a core,
and no redshift for 15 identified galaxies with $\sim$$18.5<R<21.7$ mag.

In this Paper we (i) update the original sample presented in Paper I with the new
data. These data include further radio core detections, related to that new
identifications with a host optical object (usually a galaxy) and its photometry,
spectroscopic redshifts for 4 galaxies with $\sim$$18.5<R<\sim$$19.5$ mag, and
photometric redshift estimates for the faintest identified galaxies. We also (ii)
report new VLA observations and present 4.9 GHz maps of the sample sources taken
with the C-array or D-array configurations, and derive the sources' polarisation
parameters. 

The detailed procedure and technique of the new optical and radio observations
are described in Sect.~2. The deep optical images of the newly detected host
galaxies, the long-slit optical spectra, the faint radio cores detected, and the
4.9-GHz total-intensity and polarised-intensity maps are shown in Sect.~3. The
updated sample is given in Sect.~4. The radio maps are analysed, and the physical
parameters, as well as the polarisation properties of the sample sources, are
summarised in Sect.~5. Throughout this paper we use the cosmological constants
$H_{0}$=71km\,s$^{-1}$Mpc$^{-1}$ and $\Omega_{\rm m}$=0.27. The assumption of
these constants reduce the conventional giant-size limit of 1 Mpc to about 700 kpc.
 
\section{New observations}
\subsection{Optical imaging and spectroscopy}

Deep optical imaging and $R$-band photometry of objects within three fields
comprising a radio core region of the unidentified yet sample sources
were conducted using the Asiago Astrophysical Observatory 1.8m telescope at
Cima Ekar (Italy). The CCD frames were taken through the $R$-band filter of the
Kron-Cousins $VRI$ photometric system (Cousins 1976). Exposures of the two standard
fields NGC\,2419 and NGC\,4147 (Christian et al. 1985) comprising standards up to
$R\sim 20.7$ mag were used for the magnitude calibration.

The astrometric calibration was done transforming the instrumental pixel
coordinates of stars in the investigated frame into their sky coordinates in the
DSS data base. Instrumental magnitudes of objects in the frames, reduced for bias,
dark current, and flat field with the ESO MIDAS package, were determined using
the aperture method. These instrumental magnitudes were then transformed into
Cousins $R$ magnitudes, and their errors were determined in the manner
described in Paper I.
As a result, a host galaxy was detected and its apparent $R$-band magnitude
determined for the sources: J0725+3025, J0816+3347, and J1428+3938. These
magnitudes and their errors are given in Sect. 3.1., and marked with bold-face
in column~8 of Table~3.

To detect optical spectrum and determine redshifts of the identified galaxies 
that are fainter than about 18.5 mag in the $R$-band, a large effort was made
to get observing time at optical telescopes of the 3m class. Only for 6 of the 
16 faint host galaxies, spectroscopic observations were conducted with the 3.5m
telescope of the Apache Point Observatory (Texas) equipped with low-resolution
two-sided spectrograph DIS-II and covering the spectral range of 3750--5600 \AA
(blue side) and 5500--9000 \AA (red side). A $1\farcs5$ wide slit was used,
providing the dispersion of 3.15 \AA\,pixel$^{-1}$ and the spectral resolution
of about 7 {\AA}. The wavelength calibration  was carried out using exposures
to argon/neon/helium tube and flux density calibration by short exposures of a
spectroscopic standard star close to the observed galaxy. The limited exposures
of about 30 min were taken, which was sufficient to determine a redshift for
galaxies with emission lines strong enough to be detected, but not for
galaxies only with continuum emission and absorption bands (Chy\.{z}y et al.
(2005). Consequently, it was possible to determine the redshift of the
4 of 6 observed galaxies (cf. Table~3). 

All the spectra obtained were reduced in the standard way, i.e. corrected for bias,
dark current, and flat field, cleaned of cosmic rays, combined, and calibrated
using the {\sc longslit} package of the IRAF software. One-dimensional spectra
were extracted interactively, using the observed intensity profile along the slit
to define the apertures and sky-background regions.

\subsection{VLA 4.9 GHz mapping and data reduction}

Out of 36 sample sources 20 were observed with the VLA in the C-array or D-array
at 4860 or 4885 MHz. The observing log is given in Table~1. The primary aim of
the C-array observations was to detect faint radio cores. The D-array observations
were used to map the extended lobes. For those sample sources with an angular
size not exceeding 400$\arcsec$, the field of view centred at a midpoint between
the lobes was observed using integration times from 20 min up to over 180 min,
depending on the average source's brightness. Lobes of those sources extending
more than 400$\arcsec$ were observed separately. These integration times allowed
us to reach rms noise values from about 20 $\mu$Jy\,beam$^{-1}$ to about 80
$\mu$Jy\,beam$^{-1}$. The interferometric phases were calibrated approximately
every 20 min with the phase calibrator nearest to the observed source. The sources
3C\,48, 3C\,147, and 3C\,286 were used as the primary flux density calibrators,
with the last also serving as the polarisation calibrator. 

The data were edited and calibrated using the NRAO {\sc aips} software package.
Corrections for the instrumental polarisation were determined for each antenna
using consecutive observations of the phase calibrators made at several different
hour angles. Polarisation position angles were measured relative to that of
3C286. The data were not corrected for ionospheric Faraday rotation, which
could not be significant in the $C$ band. After completing the continuum and
polarisation calibrations, the maps of the Stokes parameters $I$, $Q$ and $U$
were made and initially CLEANed with the task {\sc imagr}. Usually
several self-calibrations were performed to improve their quality. Then the maps
were corrected for the primary-beam attenuation. Because of different original
beam size, the maps were finally convolved to the common angular resolution of
$15\arcsec\times 15\arcsec$, $20\arcsec\times 15\arcsec$, or $20\arcsec\times 
20\arcsec$. 

The maps of linearly-polarised intensity and fractional polarisation were made by
combining the $Q$ and $U$ maps with the task {\sc comb}. They allow determination
of the polarised flux density at 4.9 GHz, $S_{\rm p}$, the mean fractional polarisation
$\bar{p}_{4.9}$, and the mean polarisation angle ($E$-vector) $\bar{\chi}_{4.9}$
in the lobes of the sample sources. The analysis of the errors in the total-intensity
and polarised-intensity flux densities, measured on these maps, is given in Sect.~5.2.

\begin{table}[h]
\caption{The 4.9-GHz VLA observing log}
\begin{tabular}{llll}
\hline
Array  & Project    & Observation  & Sources\\
& & date\\
\hline
D    & AJ271 & 2000,Sept30  & J1343+3758(SW lobe)\\
C    & AJ282 & 2001,July2   & J1604+3438, J1615+3826\\
     &       & 2001,July13  & J1253+4041, J1330+3850\\
D    & AM743 & 2003,Feb20   & J0912+3510\\
     &       & 2003,Feb25   & J1343+3758(NE lobe),\\
     &       &              & J1451+3357\\
D    & AM794 & 2004,June30  & J1428+3938, J1445+3051,\\
     &       &              & J1525+3345, J1554+3945,\\
     &       &              & J1555+3653, J1604+3731,\\
     &       &              & J1649+3114, J1712+3558,\\
     &       &              & J1725+3923\\
D    &       & 2004,July31  & J0720+2837, J0927+3510,\\
     &       &              & J1113+4017, J1155+4041\\
\hline
\end{tabular}
\end{table}

\section{Observational results}
\subsection{Deep optical images}

The deep optical images of the field around the sample sources J0725+3025 and
J0816+3347 are shown in Figs.~1a,\,b. In the first of these sources, the
$R$=22.56$\pm$0.05 mag galaxy perfectly coincides with the radio core
detected in the FIRST survey. The $R$=20.29$\pm$0.04 mag object identified
with the core of the second source is described in `Notes on individual sources'
in Table~3. The optical field around the source J1428+3938 and its
$R$=21.11$\pm$0.07 mag identification is shown in Fig.\,10, where the new
4.9 GHz contour map is overlaid on it. In the fields around J0725+3025 and
J1428+3938, the faint identified galaxies lie in short angular separation from
very bright foreground stars.

\begin{figure*}[t]
\resizebox{195mm}{!}{\includegraphics{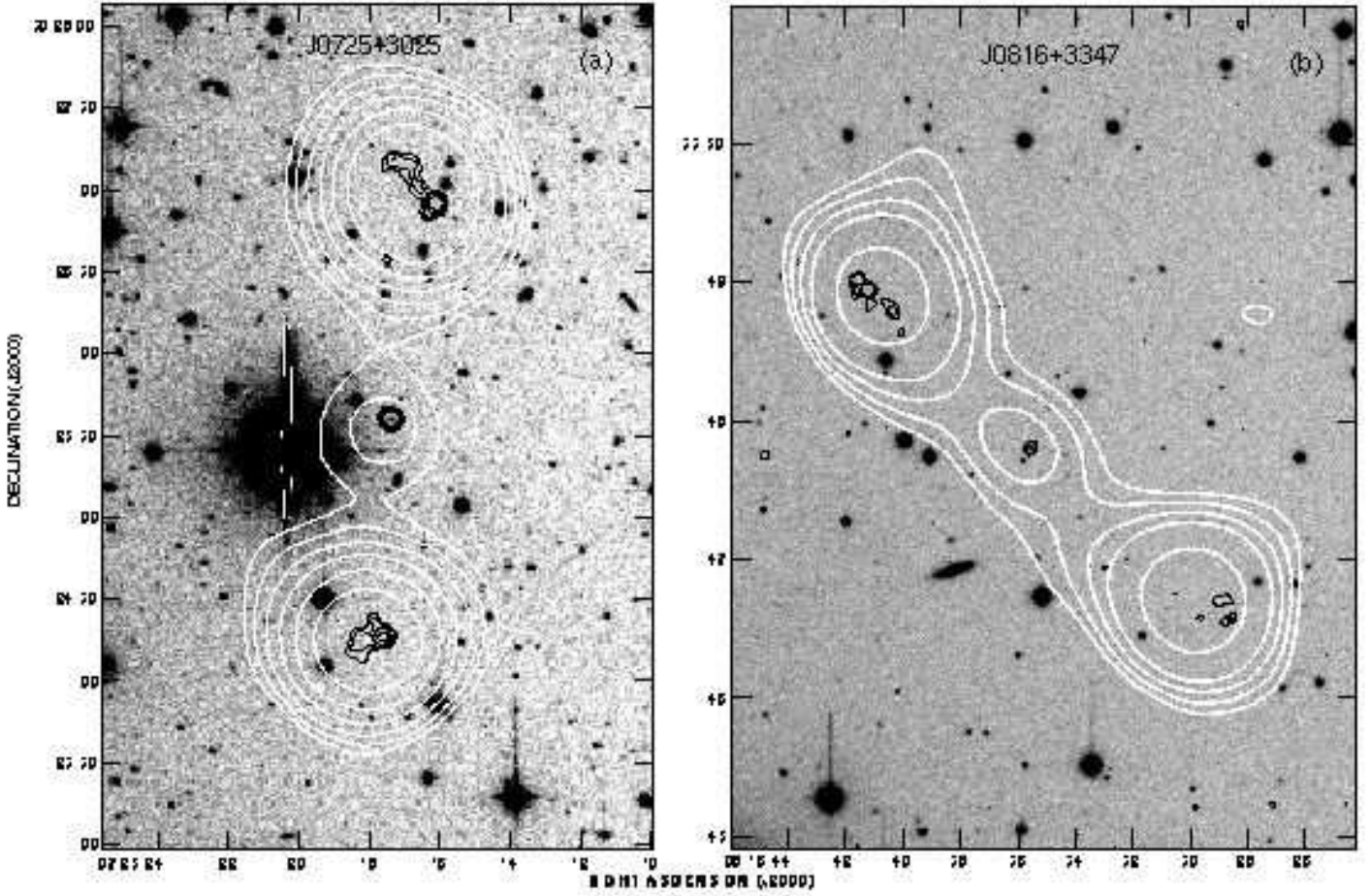}}
\caption{Deep images of the optical fields taken with the Asiago 1.8m telescope:
{\bf (a)} around the source J0725+3025 and {\bf (b)} around the source
J0816+3347. The black contours indicate the 1.4 GHz brightness distribution
observed in the FIRST survey, and white contours -- the one in the NVSS survey}
\end{figure*}

\subsection{Long-slit spectra}

The one-dimensional spectra of the observed sample galaxies are shown in
Fig.\,2. Three of four spectra exhibit prominent  emission lines:
[OII]$\lambda$3727 and [OIII]$\lambda$4959 and $\lambda$5007. The redshift
determined for the galaxy J1712+3558 is based both on the absorption lines
CaII$\lambda$3934 and $\lambda$3968, and on a trace of the line [OII]$\lambda$3727.
Preliminary spectra for the sources J1330+3850 and J1513+3841 show no emission
lines, and their crude redshift (given in Table~3) is estimated from the shape
of calibrated continuum compared by eye with the template spectrum of elliptical
galaxies given by Kennicutt (1992).

\begin{figure*}[th]
\resizebox{170mm}{!}{\includegraphics{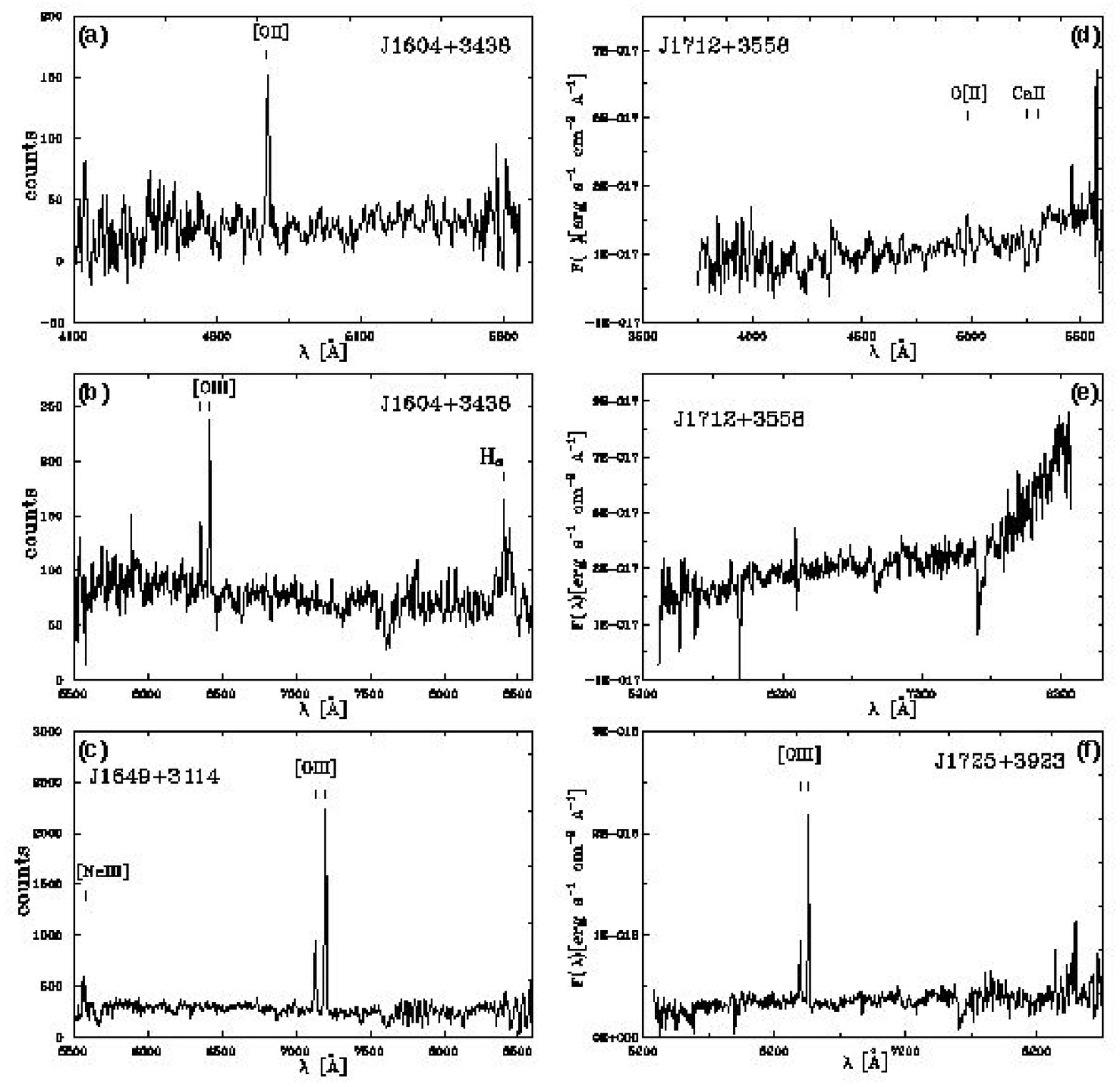}}
\caption{Low-resolution spectra of the sample galaxies taken with the 3.5m
Apache Point Observatory telescope}
\end{figure*}

\subsection{The faint radio-core detection}

\begin{table*}[t]
\caption{Data of the radio cores detected only at 4.86 GHz}
\begin{tabular*}{138mm}{@{}lccccccl}
\hline
Source    & RA(J2000) & Dec(J2000) & $S_{\rm core}$ & Deconv. size & $\Delta$\,RA & $\Delta$\,Dec & 
Notes\footnote{given in Sect.\,2}\\
&[h$\;\;\;\;$m$\;\;\;\;$s] &[$\degr\;\;\;\;\arcmin\;\;\;\;\arcsec$] & [mJy] & [$\arcsec\,\times\,\arcsec$]
& [s] & [$\arcsec$]\\
\hline
J0912+3510 & 09 12 51.95 & +35 10 14.1 & 0.24$\pm$0.04 & 1.0$\times$1.0 & +0.14 & $-$2.0 & +\\
J0927+3510 & 09 27 50.13 & +35 10 48.8 & 0.15$\pm$0.06 & 12.0$\times$$<$1.0&$-$0.45&$-$1.7\\
J1253+4041 & 12 53 12.29 & +40 41 24.5 & 1.00$\pm$0.03 & 1.3$\times$1.3 & +0.05 & +1.3\\
J1330+3050 & 13 30 36.19 & +38 50 19.8 & 0.64$\pm$0.09 & 4.8$\times$3.9 &$-$0.14 & $-$0.1\\
J1604+3438 & 16 04 45.88 & +34 38 16.1 & 1.04$\pm$0.06 & 1.2$\times$0.6 & $-$0.04 & $-$0.2 & +\\
J1615+3826 & 16 15 52.21 & +38 26 30.9 & 0.40$\pm$0.07 & $<$0.2$\times$$<$0.2&$-$0.02&$-$0.8\\
\hline
\end{tabular*}
\end{table*}

An effort was made to detect a faint radio core in distant sample sources, which
was not achieved in Paper~I, particularly for the sources unsuccessfully observed
in the years 1997--2000, i.e. J0912+3510 and J0927+3510. The sky coordinates
and the integrated flux density of the cores, newly detected in the radio maps
presented in the next subsection, are given in Table~2. The $\Delta$\,RA and
$\Delta$\,Dec give the radio-minus-optical offset with respect to the radio core. 

\subsection{4.9 GHz total-intensity and polarised-intensity maps}

The VLA 4860 or 4885 MHz images of the sample sources observed are presented in
Figs.\,3 -- 20. The total-intensity contour maps shown in panels (a) are
overlaid onto the optical field taken either from the DSS data base or our
own deep image (e.g. for J0927+3510, J1155+4029, and J1428+3938). Superimposed
are $E$-vectors (rotated by 90$\degr$) of the linearly polarised emission with
their length proportional to the polarised intensity $S_{\rm p}$. The contour
maps of linearly polarised emission with the superimposed vectors of the
fractional linear polarisation, $p_{4.9}$, are shown in panels (b). 
The full squares in polarised-intensity maps indicate the positions of the evident
hot spots. In Fig.~20 only the total-intensity map is shown for the source J1615+3826.
It is the only sample source for which we could not determine any polarisation
due to its very low surface brightness.

\begin{table*}[ht]
\caption{The updated sample. New entries are marked in bold face}
\begin{tabular*}{145mm}{@{}llrlrlrlllll}
\hline
IAU name   & FR  & $S_{\rm 1.4}$ & $\alpha_{\rm hf}$ & $S_{\rm core}$ & LAS &$AR$& 
opt. & $R$ & $z$ & Ref. & Notes\\
(J2000)    & type& [mJy]  &      & [mJy] & ["] & & id.  & [mag]\\
\hline
J0720+2837 & II  &  46 &{\bf 0.79} & 1.03 & 376 & 8.4 & G   & 17.75 & 0.2705 & 1,6\\
J0725+3025 & II  &  35 & 1.4  & 2.72 & 175 & 6.0 &{\bf G}&{\bf 22.56}&{\bf (0.70)} & 7\\
J0816+3347 & IID &  40 &{\bf 0.80}  & 1.26 & 210 & 5.8 & ?   &{\bf 20.29}&{\bf (0.42)} & 7 & +\\
J0912+3510 & II  & 161 &{\bf 0.90} &{\bf 0.24}& 375 & 9.4 & G& 19.35 & 0.2489 & 1,6 & +\\
J0927+3510 & II  &  96 &{\bf 0.89} &{\bf 0.15}& 345 &11.5  & G& 21.7  &{\bf (0.55)} & 6,7 & +\\
J1011+3111 & II  &  71 & 0.7  & 1.13 & 285 & 8.1 & G   & 21.2  &{\bf (0.50)} & 6,7\\
J1113+4017 & I/II& 249 & 0.75 &$<$15.2\,\,\,& 646 & 11.1 & G   & 14.56 & 0.0745 & 1,5\\
J1155+4029 & II  & 323 & 1.00 & 13.7\,\,\,& 229  &13.5 & G & 21.5  &{\bf (0.53)} & 6,7 & +\\
J1200+3449 & II  & 227 & 1.01 & 0.36 & 147  & 8.6 & G   & 21.2  &{\bf (0.50)} & 6,7\\
J1253+4041 & IID &  52 &{\bf 0.6}  &{\bf 1.00}&275 & 5.7 & G& 17.27 & 0.2302 & 1,6 & +\\
J1254+2933 & II  &  63 & 0.6  & 0.74 & 295  &12.8 & G   & 20.3  &{\bf (0.42)} & 6,7\\
J1330+3850 & IID &  31 & 0.77  &{\bf 0.64}& 280 & 8.0 & G& 19.3  &{\bf 0.63?} & 2,6 & +\\
J1343+3758 & II  & 140 & 0.88 & 1.09 & 678  &12.1 & G   & 17.94 & 0.2267 & 1,6 &+\\
J1344+4028 & I/II& 209 & 0.76 &$<$22.0\,\,\,& 450 & 6.8 & G & 14.92 & 0.0748 & 1,6 & +\\
J1345+3952 & II  & 170 &{\bf 1.05} &$<$1.0\,\,\,& 192  & 5.1 & G   & 15.80 & 0.1611 & 1,6\\
J1355+2923 & II  & 140 &{\bf 1.02} & 1.86 & 263  & 4.4 & G   & 20.4  &{\bf (0.43)} & 6,7\\
J1428+2918 & II  & 448 & 1.04 & 12.8\,\,\,& 905 & 7.9 & G   & 13.0  & 0.0870 & 5\\
J1428+3938 & IID &  90 &{\bf 1.23} & 3.46 & 269 & 9.6 &{\bf G}&{\bf 21.11} &{\bf (0.50)} & 7\\
J1445+3051 & II  &  97 & 0.90 & 10.9\,\,\,& 300  & 8.1 & G   & 19.0  & 0.42  & 4\\
J1451+3357 & II  & 142 & 0.91 & 3.89 & 245 & 4.4 & G   & 18.7  & 0.3251 & 6\\
J1453+3309 & II  & 455 & 0.94 & 3.84 & 320 & 7.6 & G   & 18.3  & 0.249  & 5\\
J1512+3050 & II  & 107 & 0.64 & 5.4* & 240 & 5.3 & G   & 15.98 & 0.0895 & 1,6\\
J1513+3841 & IID &  23 &{\bf 0.7} &$<$1.0\,\,\,& 267 & 4.8 &{\bf G}&{\bf 19.4}&{\bf 0.52?} & 2,7 & +\\
J1525+3345 & IID &  51 &{\bf 0.83} & 2.41 & 215 & 5.4 & G   & 20.9  &{\bf (0.47)} & 6,7\\
J1526+3956 & II  &  62 & 1.0  &  ?\,\,\,& 536 & ? &... & ....  & ....   & \\
J1554+3945 & II  &  74 &{\bf 0.82} & 0.84 & 220 & 8.2 & G   & 19.5  &{\bf (0.35)} & 6,7\\
J1555+3653 & II  & 106 &{\bf 0.94} & 15.5\,\,\,& 349  & 8.1 & G & 18.56 & 0.2472 & 1,6\\
J1604+3438 & II  & 146 & 0.95 &{\bf 1.04}& 200 & 5.0 &G& 18.8  &{\bf 0.2817}& 1,2 & +\\
J1604+3731 & II  & 122 & 1.03 & 2.74 & 182  & 5.2 & G?  & ....  & 0.814  & 3 & +\\
J1615+3826 & IID &  34 &{\bf 0.77}&{\bf 0.40}& 264 & 6.9 & G& 17.59 & 0.1853 & 1,6 & +\\
J1632+3433 & II  &  28 & 0.9  & 3.9* & 180 & 7.2 &... & ....  & ....   & \\
J1635+3608 & I/II& 100 & 0.66 & 3.8* & 320 &10.7 & G   & 17.30 & 0.1655 & 1,6\\
J1649+3114 & II  & 153 &{\bf0.82} & 2.22 & 209 & 9.1 & G   & 19.62 &{\bf 0.4373}& 1,2\\
J1651+3209 & I/II&  68 & 0.6  & 28.1*& 430 & ?&... & ....  & ....   & \\
J1712+3558 & II  &  87 &{\bf 0.76} & 0.55 & 211 & 7.5 & G & 19.1  &{\bf 0.3357}& 1,2\\
J1725+3923 & II  &  88 &{\bf 0.91} & 1.08 & 288 & 6.9 & G & 18.76 &{\bf 0.2898}& 1,2\\
\hline  
\end{tabular*}
\vspace{2mm}
\begin{tabular*}{135mm}{@{}rlrl}
{\bf References}\\
(1)& Digitized Sky Survey (DSS) &  (5)& Schoenmakers et al. (2001)\\
(2)& Chy\.{z}y et al. (2005)    &  (6)& Paper I\\
(3)& Cotter et al. (1996)       &  (7)& this paper\\
(4)& Hook et al. (1998)
\end{tabular*}
\end{table*}

\section{The updated sample}

The update of the sample is presented in Table~3 where the updated entries
and new data are marked in bold face. The consecutive columns give:

{\sl Column 1:} IAU name at epoch J2000.

{\sl Column 2:} Fanaroff-Riley morphological type. `D' indicates a diffuse
morphology of the radio lobes without any detectable hot spot(s).

{\sl Column 3:} 1.4 GHz total flux density in mJy.

{\sl Column 4:} Radio spectral index between 325  and 4860 MHz.

{\sl Column 5:} Flux density of the radio core determined at 5 GHz. The values
available only at 1.4 GHz are marked with asterisks.
 
{\sl Column 6:} Largest angular size in arcsec.

{\sl Column 7:} Axial ratio of the source measured as the ratio between the largest
angular size (cf. column~6) and the average of the full width of its two
lobes. The latter is determined as the deconvolved width of the transversal
cross-section through the lobes measured between 3$\sigma$ total-intensity
contours half-way between the core and the outer edge of the source.

{\sl Column 8:} Optical identification.

{\sl Column 9:} Apparent $R$-band magnitude of the identified host optical
object taken from the reference given in Col.\,10.

{\sl Column 10:} Redshift of the host object (galaxy). The value in parenthesis
is the photometric redshift estimate based on the assumption that the host optical
counterpart is a galaxy with $M_{\rm R}$=$-23\fm 65$ (cf. Paper I).

{\sl Column 11:} References to the optical magnitude and redshift.

{\sl Column 12:} Notes about individual sample sources (given below).

\vspace{2mm}
\noindent
{\bf Notes on individual sources in Table 3:}

{\sl J0816+3347:} The radio core coincides with a very blue object in the POSS1
($O-E$=$-0\fm 0.4$; $E=18\fm 76$). Our photometry in March 2000 gave $R=20\fm 7\pm 0\fm 2$,
and $R=20\fm 29\pm 0\fm 04$ in Dec. 2004. The blue colour and a probable optical
variability may suggest a QSO, however the diffused radio lobes and the low-brightness
bridge evident in the NVSS map (cf. Fig.~1b) are typical for radio galaxies.

{\sl J0912+3510:} The detected compact core is 2.6$\arcsec$ from the identified galaxy
(cf. Table~2). This identification and the measured redshift of 0.2489 may not be certain
due to a poor coincidence with the radio core and the fact that the object visible
in the DSS field (marked with "A" in Fig.~4a) appears to be a close pair of
galaxies that can be unrelated. Figure~4c shows the deep direct image taken with the
McDonald 2.1m telescope. The two galaxies are separated by about $3\farcs5$, while the
optical spectrum was measured through the $2\arcsec$ slit (cf. Paper I), therefore
there is some doubt as to which galaxy was actually measured due to a limited
telescope-pointing accuracy of about 2$\arcsec$--3$\arcsec$. 

{\sl J0927+3510:} Possibly of a double-double morphology. The two inner components
(cf. Fig.~5a) surround the faintest core detected among the sample sources. Their
brightness peaks at the positions of $09^{\rm h}27^{\rm m}45\fs97$;
$+35\degr 11\arcmin 00\farcs8$ and $09^{\rm h}27^{\rm m}53\fs56$;
$+35\degr 10\arcmin 41\farcs9$ determined from the VLA 8.46 GHz map. The outer, much
brighter lobes are terminated by hot spots clearly visible in that map shown in
Paper~I.

{\sl J1155+4029:} Extremely large flux-density ratio between the two lobes measured at
four frequencies [325 MHz: WENSS (Rengelink et al. 1997); 408: B3 (Ficarra et al. 1985);
1400 MHz: NVSS (Condon et al. 1998); and 4860 MHz: this paper] and the characteristic
inverse ratio of the lobes' separation may suggest a less than 90$\degr$ inclination
angle of this source to the observer line.

{\sl J1253+4041:} The detected radio core (cf. Table~2) confirms the optical
identification given in Paper~I.

{\sl J1330+3850:} The detected radio core (cf. Table~2) coincides perfectly with the
$R$$\approx$19.3 mag galaxy, the certain host galaxy of this large source with strongly
diffused lobes. Its redshift as suggested by Chy\.{z}y et al. (2005) remains uncertain
because no evident emission lines were detected in the optical spectrum.

{\sl J1343+3758:} This source has been analysed in detail by Jamrozy et al. (2005)
where its synchrotron and dynamical ages are determined, and a number of physical
parameters derived.

{\sl J1344+4027:} This extended double source of FRI/II-type morphology is badly
confused by a strong background FRII-type source.

{\sl J1513+3841:} The redshift is uncertain for the same reason as that in J1330+3850.

{\sl J1604+3438:} Possibly of a double-double morphology.
 
{\sl J1604+3731:} The 613 MH GMRT map of this source has recently been published by
Konar et al. (2004).

{\sl J1615+3826:} The radio core detected (cf. Table~2) confirms the optical
identification given in Paper~I.

\section{Analysis of radio maps}
\subsection{Physical parameters derived from the data}

The NVSS, FIRST, and our VLA maps are used to determine the basic physical
parameters of the sample sources. These parameters are given in Table~4; their
value, which is less certain because of the preliminary photometric redshift
estimate, is marked with asterisk. 

\begin{table*}[t]
\caption{Physical parameters of the sample sources}
\begin{tabular*}{150mm}{@{}lrcrrrrcccc}
\hline
Source   & $D\;\;$ & log$P_{\rm 1.4}$ & log$P^{\rm core}_{\rm 5}$ & $c_{\rm p}\;\;$
& ${\cal P}_{CN}$ & $B_{\rm eq}$ & $u_{\rm eq}10^{-14}$ & $U\,10^{52}$ & $
\frac{B_{\rm iC}}{B_{\rm eq}}$\\
         & [kpc]&[W\,Hz$^{-1}$]& [W\,Hz$^{-1}$]&&& [nT] & [J\,m$^{-3}$] & [J]\\
\hline
J0720+2837 &$1542\;\,$& 24.99 & 23.26 &  0.018 &  0.58 & 0.092 & 0.79 & 0.95 & 5.57\\
J0725+3025 & 1246*& 25.9* & 24.5* &  0.045 &  2.88 & 0.190 & 3.34 & 4.09 & 4.84\\
J0816+3347 & 1155*& 25.4* & 23.7* &  0.024 &  0.91 & 0.155 & 2.24 & 2.36 & 4.13\\
J0912+3510 &$1444\;\,$ & 25.46 & 22.55 &  0.001 &  0.06 & 0.162 & 2.43 & 1.92 & 3.06\\
J0927+3510 & 2200*& 26.0* & 23.1* &$<$0.001&$<$0.09& 0.153 & 2.18 & 4.09 & 4.98\\
J1011+3111 & 1740*& 25.8* & 23.9* &  0.012 &  0.83 & 0.134 & 1.67 & 3.02 & 5.33\\
J1113+4017 & $886\;\,$ & 24.49 &$<$23.2\,\,\,&$<$0.058&  1.01 & 0.113 & 1.18 & 3.05 & 3.25\\
J1155+4029 & 1440*& 26.5* & 24.9* &  0.028 &  3.16 & 0.487 & 22.0 & 8.30 & 1.53\\
J1200+3449 &  890*& 26.3* & 23.3* &  0.001 &  0.10 & 0.473 & 20.8 & 4.57 & 1.51\\
J1253+4041 & $996\;\,$ & 24.86 & 23.09 &  0.017 &  0.47 & 0.098 & 0.89 & 0.62 & 4.93\\
J1254+2933 & 1630*& 25.5* & 23.5* &  0.009 &  0.50 & 0.161 & 2.42 & 1.46 & 3.97\\
J1330+3850 &$1910\;\,$ & 25.67 & 23.80 &  0.014 &  0.79 & 0.148 & 2.03 & 5.10 & 5.71\\
J1343+3758 &$2427\;\,$ & 25.30 & 23.12 &  0.007 &  0.28 & 0.103 & 0.99 & 2.23 & 4.63\\
J1344+4028 & $630\;\,$ & 24.42 &$<$23.42&$<$0.100 &     & 0.128 & 1.51 & 0.19 & 2.88\\
J1345+3952 & $520\;\,$ & 25.06 &$<$22.8\,\,\,&$<$0.005&$<$0.18& 0.196 & 3.57 & 0.45 & 2.19\\
J1355+2923 &$1468\;\,$ & 25.97 & 23.94 &  0.009 &  0.72 & 0.152 & 2.15 & 1.98 & 4.27\\
J1428+2918 &$1412\;\,$ & 24.89 & 23.31 &  0.026 &  0.75 & 0.089 & 0.73 & 0.77 & 4.23\\
J1428+3938 & 1630*& 26.0* & 24.4* &  0.023 &  1.99 & 0.272 & 6.84 & 7.46 & 2.64\\
J1445+3051 &$1653\;\,$ & 25.77 & 24.69 &  0.083 &  5.35 & 0.140 & 1.82 & 2.88 & 4.58\\
J1451+3357 &$1147\;\,$ & 25.68 & 24.00 &  0.021 &  1.24 & 0.143 & 1.89 & 3.44 & 3.92\\
J1453+3309 &$1238\;\,$ & 25.92 & 23.75 &  0.007 &  0.51 & 0.199 & 3.68 & 2.78 & 2.49\\
J1512+3050 & $396\;\,$ & 24.32 & 23.00 &  0.048 &  0.81 & 0.129 & 1.54 & 0.08 & 2.93\\
J1513+3841 &$1655\;\,$ & 25.33 &$<$23.8\,\,\,&$<$0.032&$<$1.26& 0.096 & 0.85 & 3.90 & 7.69\\
J1525+3345 & 1260*& 25.6* & 24.1* &  0.032 &  1.74 & 0.145 & 1.95 & 3.14 & 4.74\\
J1554+3945 & 1080*& 25.5* & 23.4* &  0.009 &  0.41 & 0.173 & 2.76 & 1.22 & 3.36\\
J1555+3653 &$1335\;\,$ & 25.28 & 24.35 &  0.117 &  4.81 & 0.139 & 1.80 & 1.50 & 3.55\\
J1604+3438 & $845\;\,$ & 25.55 & 23.30 &  0.006 &  0.29 & 0.182 & 3.07 & 1.71 & 2.87\\
J1604+3731 &$1376\;\,$ & 26.59 & 24.67 &  0.012 &  1.64 & 0.272 & 6.85 & 15.2 & 3.85\\
J1615+3826 & $805\;\,$ & 24.49 & 22.50 &  0.010 &  0.20 & 0.097 & 0.87 & 0.22 & 4.62\\
J1635+3608 & $895\;\,$ & 24.83 & 23.38 &  0.036 &  0.96 & 0.180 & 3.02 & 0.36 & 2.48\\
J1649+3114 &$1179\;\,$ & 26.00 & 24.03 &  0.011 &  0.85 & 0.278 & 7.19 & 3.30 & 2.36\\
J1712+3558 &$1013\;\,$ & 25.48 & 23.18 &  0.005 &  0.25 & 0.184 & 3.14 & 1.33 & 3.08\\
J1725+3923 &$1244\;\,$ & 25.35 & 23.33 &  0.010 &  0.42 & 0.132 & 1.62 & 1.53 & 4.01\\
\hline
\end{tabular*}
\end{table*}
 
After the source name (column~1), the subsequent columns are:

{\sl Column 2:} Projected linear size calculated with $H_{0}$=71\,km\,s$^{-1}$Mpc$^{-1}$
and $\Omega_{\rm m}$=0.27. These values of cosmological constants cause a reduction in
size for most sample sources with respect to their value of $D>1$ Mpc determined with
$H_{0}$=50\,km\,s$^{-1}$Mpc$^{-1}$, as used in most previously published papers.

{\sl Column 3:} Logarithm of total radio power at the emitted frequency of 1.4 GHz
calculated using the cosmological constants as in Col.~2.

{\sl Column 4:} Logarithm of the core power at the emitted frequency of 4.9 GHz.

{\sl Column 5:} Core prominence defined as $c_{\rm p}=P^{\rm core}_{5}/(P_{1.4}
-P^{\rm core}_{5})$.

{\sl Column 6:} Source axis orientation indicator defined as the ratio between the
observed core luminosity, $P_{5}^{\rm core}$, and its value expected from the
statistical relation of Giovannini et al. (2001), scaled here from their original
408 MHz total power to the total power at 1.4 GHz assuming a spectral index of 0.75,
and transformed to the cosmological constants adopted in this Paper: 
${\cal P}_{CN}=P^{\rm core}/(P_{1.4}^{0.60}+8.5)$.

{\sl Column 7:} Average magnetic field strength, $B_{\rm eq}$, calculated using the
formula of Miley (1980), i.e. under assumption of energy equipartition, a cylindrical
geometry of the extended emission with the base diameter equal to the average width of
the lobes, usually measured half-way between the core and the brightest regions, using
the prescription of Leahy \& Williams (1984), a filling factor of unity, and the
equipartition of kinetic energy between relativistic electrons and protons. The total
radio luminosity is integrated from 10 MHz to 100 GHz.

{\sl Column 8:} Minimum energy density, $u_{\rm eq}$, calculated under the same
assumptions.

{\sl Column 9:} Total energy contained in the source, i.e. the minimum energy density
times the volume of the source's cocoon.

{\sl Column 10:} Ratio of the equivalent magnetic field strength of the microwave
background radiation, $B_{\rm iC}$=0.318(1+$z$)$^{2}$[nT], to the equipartition magnetic
field $B_{\rm eq}$. 

\subsection{Fractional polarisation, depolarisation, and rotation measures}

\begin{table*}[t]
\caption{Polarisation data for the sample sources}
\begin{tabular*}{179mm}{@{}llcc}
\hline
Source & &\hspace{25mm} Lobe closer to the core &\hspace{37mm} Lobe farther to the core
\end{tabular*}
\begin{tabular*}{179mm}{@{}llrrcrlrrcr}
&lobe & $\bar{p}_{4.9}$ & $\bar{\chi}_{4.9}$ & DP & RM &
lobe & $\bar{p}_{4.9}$ & $\bar{\chi}_{4.9}$ & DP & RM\\   
           & & $\pm\delta\bar{p}_{4.9}$[\%] & $\pm\sigma_{4.9}$[$\degr$] & & [rad\,m$^{-2}$]& &
               $\pm\delta\bar{p}_{4.9}$[\%] & $\pm\sigma_{4.9}$[$\degr$] & & [rad\,m$^{-2}$]\\
\hline
J0720+2837 & SW& 4.7$\pm$2.5 & +36$\pm$27 & &
           & NE& 4.4$\pm$2.1 &$-$44$\pm$34\\
J0912+3510 & N &18.5$\pm$2.6 & +44$\pm$8 & 1.11$\pm$0.19 & +15$\pm$7
           & S &17.9$\pm$1.6 & +40$\pm$9 & 0.90$\pm$0.11 & +13$\pm$7\\
J0927+3510 & E &12.1$\pm$1.6 &$-$6$\pm$6 & 1.14$\pm$0.21 & +6$\pm$9
           & W &16.1$\pm$2.3 & +11$\pm$8 &  1.34$\pm$0.26 & +12$\pm$8\\
J1113+4017 & NE&13.7$\pm$1.0 &$-$27$\pm$5&  1.49$\pm$0.15 & +11$\pm$6
           & SW&19.5$\pm$1.2 &$-$33$\pm$6&  1.03$\pm$0.12 & +6$\pm$5\\
J1155+4029 & N & 3.0$\pm$0.4 &$-$36$\pm$16& 1.05$\pm$0.17 &$-$1$\pm$7
           & S & 8.2$\pm$1.9 &$-$49$\pm$11& \\
J1343+3758 & N &15.9$\pm$1.4 &$-$1$\pm$9 &  1.15$\pm$0.21 & +2$\pm$5
           & S & 7.1$\pm$0.7 &$-$5$\pm$7 &  1.87$\pm$0.29 &$-$0$\pm$8\\
J1428+3938 & S &14.7$\pm$2.6 &$-$42$\pm$13& 1.29$\pm$0.25 &$-$3$\pm$8
           & N &24.0$\pm$4.8 &$-$46$\pm$16& 0.83$\pm$0.22 &$-$2$\pm$10\\
J1445+3051 & S & 5.7$\pm$1.3 & +24$\pm$13&  &
           & N &11.7$\pm$2.8 & +55$\pm$11&  \\
J1451+3357 & SE& 6.5$\pm$1.0 &$-$32$\pm$7 & 0.96$\pm$0.26 & +2$\pm$9
           & NW&13.0$\pm$1.2 & +12$\pm$7  & 0.96$\pm$0.15 & +3$\pm$7\\
J1525+3345 & N & 5.3$\pm$5.0 & +14$\pm$13 & &
           & S & 5.8$\pm$2.3 &$-$40$\pm$25& \\
J1554+3945 & S &12.7$\pm$2.0 & +60$\pm$15 & 1.02$\pm$0.25 &$-$8$\pm$6
           & N &18.9$\pm$2.2 & +67$\pm$5  & 0.95$\pm$0.18 &$-$2$\pm$6\\
J1555+3653 & N & 6.8$\pm$2.6 &$-$8$\pm$24 & &
           & S &10.0$\pm$1.2 &$-$38$\pm$24& 0.72$\pm$0.24 & +2$\pm$7\\
J1604+3438 & E & 7.9$\pm$2.0 &$-$18$\pm$10& &
           & W & 8.1$\pm$2.2 & +38$\pm$18 & \\
J1604+3731 & S & 8.5$\pm$2.9 &$-$65$\pm$18& &
           & N & 9.3$\pm$3.6 & +35$\pm$17 & \\
J1649+3114 & N &16.9$\pm$0.8 &$-$3$\pm$11 & 1.04$\pm$0.08 & +18$\pm$7
           & S & 3.5$\pm$1.7 &$-$56$\pm$18& \\
J1712+3558 & S & 5.2$\pm$1.1 &$-$8$\pm$30 & 1.11$\pm$0.32 & +13$\pm$7
           & N &32.6$\pm$5.9 & +45$\pm$9  & 0.78$\pm$0.23 & +14$\pm$8\\
J1725+3923 & S & 6.4$\pm$2.0 &$-$15$\pm$17& &
           & N & 7.6$\pm$1.6 &$-$17$\pm$16& 1.75$\pm$0.49 &$-$8$\pm$11\\
\hline
\end{tabular*}
\end{table*}

To determine the crude values of depolarisation and rotation measures for the
sample sources whose 4.9 GHz maps are shown in Figs.~3 -- 19, we use
the commonly available 1.4 GHz NVSS maps of the Stokes $I, Q, U$ parameters.
For each of those sources, the areas comprising the detected emission of
their two lobes were determined on the Stokes $I$ map, and pixels outside the
lobe area with the total intensity flux less than 3$\sigma_{\rm I}$ were blanked.
Because NVSS maps have a lower angular resolution than our maps, we first
convolved the original 4.9 GHz VLA $I, Q, U$ maps with the NVSS beam of
45$\arcsec \times 45\arcsec$. The 1.4 and 4.9 GHz maps were centred to the same
sky position with the task {\sc hgeom}, and for the sample sources with the compact
core detected at both frequencies (at 1.4 GHz: in the FIRST), the alignment was
also checked using the core position. Next, for each of the  analysed sources,
a rectangular area (box) was chosen around its two lobes, and the integrated value
of $S^{\prime}_{\rm p}$ and the mean polarisation angle $\bar{\chi}^{\prime}$ were
measured within each box using the maps of fractional polarisation and polarisation
angle, again made with the task {\sc comb}.

The mean values of the fractional polarisation for each lobe is determined as

\[\bar{p}^{\prime}_{k}=(S^{\prime}_{\rm p})_{k}/(S^{\prime}_{\rm I})_{k},\]  

\noindent
and the associated position angle is $\bar{\chi}^{\prime}_{k}$ for $k$=1.4, 4.9 GHz.
Note that the {\sl primed} symbols indicate parameters whose values are determined
from the convolved maps. We define the depolarisation measure as the ratio

\[DP=\bar{p}^{\prime}_{1.4}/\bar{p}^{\prime}_{4.9}\]
 
\noindent
although this frequently used definition implies an increasing depolarisation
if the $DP$ value decreases. According to the simple models of the `Faraday-thin'
screens (e.g. Laing 1984), the resulting low depolarisation of the sample members
($DP\gg 0.2$; cf. Table~5) implies that the difference between the rotation angles
$\bar{\chi}^{\prime}_{1.4}-\bar{\chi}^{\prime}_{4.9}$ is always less than 90$\degr$.
This allows us to calculate the Faraday rotation measure necessary to determine
the intrinsic polarisation angle in the sample sources. The two-point rotation
measure is calculated as the ratio

\[RM=(\bar{\chi}^{\prime}_{1.4}-\bar{\chi}^{\prime}_{4.9})/(\lambda^{2}_{1.4}-
\lambda^{2}_{4.9}).\]

\noindent
The error in total flux density is calculated as

\[\delta S^{\prime}_{\rm I}=(\sigma^{2}_{\rm I}\,A_{\rm int})^{1/2},\]

\noindent
where $\sigma_{\rm I}$ is the rms noise of the total-intensity map, and $A_{\rm int}$
is the integration area in units of the beam solid angle. The value of rms noise for
1.4 GHz, $\sigma_{\rm I}$=0.45 mJy\,beam$^{-1}$, is taken from Condon et al. (1998),
while its value for 4.9 GHz is measured in the particular VLA map.
The errors in polarised flux density and fractional polarisation are calculated as

\[\delta S^{\prime}_{\rm p}=\frac
{\left[(S^{\prime}_{Q}\cdot\delta S^{\prime}_{Q})^{2}+(S^{\prime}_{U}\cdot\delta
S^{\prime}_{U})^{2}\right]^{1/2}}{S^{\prime}_{\rm p}},\]

\noindent
where $\delta S^{\prime}_{i}=(\sigma^{2}_{i} A_{\rm int})^{1/2}$ for $i=Q,U$. 
The rms noise in polarised flux density at 1.4 GHz, $\sigma_{i}=0.29$ mJy\,beam$^{-1}$,
again is taken from Condon et al., and its value for 4.9 GHz is measured
in the particular VLA $Q$ and $U$ maps, and

\[\delta\bar{p}_{4.9}=\left[(\delta S^{\prime}_{\rm p}/S^{\prime}_{\rm I})^{2}+
(S^{\prime}_{\rm p}\sigma_{\rm I}/S^{\prime\,2}_{\rm I})^{2}\right]^{1/2}.\]

\noindent
The dispersion of the polarisation angle, $\sigma_{4.9}$, is determined from the
original (not convolved) polarisation angle maps by

\[\sigma_{4.9}=\left[\sum_{i}(\chi_{i}-\bar{\chi}_{4.9})^{2}/A_{\rm int}\right]^{1/2},\]

\noindent
where the sum is taken over map pixels. A similar formula is used
to determine the dispersion of the polarisation angles, $\sigma^{\prime}_{1.4}$ and
$\sigma^{\prime}_{4.9}$, which are measured in the convolved polarisation angle maps. 

The polarisation parameters determined here for the lobe closer to the radio core, and
farther off the core of the analysed sources are given in Table~5 as follows:

{\sl Column 1:} IAU name at epoch J2000.

{\sl Columns 2 and 7:} Lobe designation.

{\sl Columns 3 and 8:} Fractional polarisation at 4.9 GHz measured on the original
map with its error.

{\sl Columns 4 and 9:} Mean polarisation angle at 4.9 GHz measured on the
original map with its dispersion.

{\sl Columns 5 and 10:} Depolarisation measure with error.

{\sl Columns 6 and 11:} Rotation measure with error in rad/m$^{2}$.

\vspace{3mm}
The data in Table~5 show that depolarisation of the lobe emission in the
sample sources, in our case the giant radio galaxies, is extremely low.
Though some $DP$ values are formally much greater than unity, their uncertainty
is also high due to usually very low flux densities in the polarised-intensity
maps. The median $DP$ value for 20 lobes is 1.04$\pm$0.05; i.e. it is close
to unity. Also, the $RM$ values suggest that the Faraday rotations of the
polarisation planes are mild.

The maps in Figs.~3 -- 19  show that the dominant position
angles of the magnetic fields in the lobes are close to those of the main axes
of the sample sources. The only exception to this behaviour can be seen in the
sources J0912+3510 and J1445+3051, where the angle between the mean direction
of magnetic fields and the source's axis is larger than 45$\degr$. In the
sources with twisted lobes (e.g. J1113+4017; J1451+3357), the direction, i.e.
the magnetic field projected vectors, seems to follow bends in the lobe structures.
In most of our maps, the hot spots are not resolved from the lobes, and their
emission is strongly depolarised in the beam. However, if they are resolved out
(e.g. in J1343+3758; J1712+3558), the mean direction of the magnetic fields tends
to be almost transversal to the source's axes.

\section{Comparison with other samples}

In this section we compare some statistical characteristics of our sample with
relevant characteristics that can be derived for other samples of giant radio
sources. The main characteristics are: distributions of the redshift, projected
linear size, and monochromatic radio power. A simple one-figure presentation of
these distributions is the radio power vs. intrinsic size relation ($P-D$ diagram).
In Fig.~21 we plot this diagram for the sources selected from the samples of Lara
et al. (2001), Schoenmakers et al. (2001), Machalski et al. (2001), and Saripalli
et al. (2005), i.e. limited to the sample members of FRII-type or FRI/II-type morphology
and having spectroscopic redshift, or its reliable photometric estimate. We determine
the radio power at the frequency of 1.4 GHz using flux densities published in the
original Lara et al.'s and Schoenmakers et al.'s papers (mostly NVSS flux densities),
and the ones in Table~3 for our sample. For the Saripalli et al.'s sample, the power
is calculated by transforming original 843-MHz flux densities into 1.4 GHz ones,
assuming a mean spectral index $\alpha=-0.8$.

Since our sample was selected in the same way as the samples of Lara et al. and
Schoenmakers et al., the selection effects are very similar to those described and
discussed in the above papers. The only differences were values of the selection
limits for the minimum angular size and minimum of 1.4 GHz flux density. The
Saripalli et al.'s sample, though selected at another observing
frequency, has a limiting flux density comparable to those for the above two
samples. The dashed curved line in Fig.~21 indicates the position of a source with
the total flux density of 30 mJy at 1.4 GHz and the angular size of 6 arc min, as
a function of the source's redshift. An area below this curve is the part of the
diagram in which sources cannot be selected due to the sensitivity limit of the
finding surveys (WENSS, NVSS, SUMSS). The largest giant sources known, e.g.
B2147+816, SGRS J0331-7710, and J1343+3758 (in the diagram), as well as 3C236
(out of the diagram), lie within the above area due to their angular sizes
being significantly larger than 6 arc min.

The statistical characteristics of the four samples are given in Table~6. They
are limited to their members with FRII or FRI/II morphology, with spectroscopic
redshift or its photometric estimate, and to those larger than 700 kpc. The
consecutive columns give:

{\sl Column 1:} The sample.

{\sl Column 2:} Sky area of the sample in sr.

{\sl Column 3:} Number of sources fulfilling the above restrictions.

{\sl Column 4:} Angular size limit in arc min.

{\sl Column 5:} Approximate 1.4-GHz flux density limit in mJy.

{\sl Column 6:}  Median redshift.

{\sl Column 7:} Redshift mean deviation.

{\sl Column 8:} Median linear size in Mpc.

{\sl Column 9:} Size mean deviation in Mpc.

{\sl Column 10:} Median logarithm of 1.4-GHz power in W\,Hz$^{-1}$.

{\sl Column 11:} Logarithm of power mean deviation in W\,Hz$^{-1}$.

\begin{table*}[t]
\caption{Comparison of the samples; the median and mean deviation (MD) values in
distributions of redshift, linear size, and 1.4-GHz power}
\begin{tabular*}{175mm}{@{}lcccrllcccc}
\hline
Sample & A & N & $\Theta_{min}$& $S_{min}$& Median z & MD$_{z}$ & Median D & MD$_{D}$ &
Median log$P_{1.4}$ & MD$_{P}$\\
  & [sr] & & [$\arcmin$] & [mJy] & & & [Mpc] & [Mpc] & [W\,Hz$^{-1}$] &[W\,Hz$^{-1}$]\\
\hline
Schoenmakers et al. & -- & 22 & 5 & $\sim$100 & 0.16$\pm$0.02 & 0.06 & 1.14$\pm$0.10 & 0.32 &
25.18$\pm$0.16 & 0.47\\
Lara et al.        & 0.84 & 29 & 4 & 100 & 0.21$\pm$0.02 & 0.11 & 1.07$\pm$0.08 & 0.64 &
25.59$\pm$0.15 & 0.42\\
Saripalli et al.   & 0.64 & 15 & 5 & $\sim$100& 0.25$\pm$0.03 & 0.08 & 1.52$\pm$0.21 & 0.52 &
25.61$\pm$0.13 & 0.38\\
our          & 0.47 & 31 & 3 & 30 & 0.38$\pm$0.07 & 0.14 & 1.30$\pm$0.09 & 0.30 &
25.52$\pm$0.10 & 0.40\\
\hline 
\end{tabular*}
\end{table*}

The data in Table~6 imply that:

-- All the four samples probe a same part of the population of extragalactic
radio sources, especially as regards the range of radio power.  There is 
practically no difference in the median deviations of 1.4-GHz power, though the
median power in the Schoenmakers et al.'s sample is about 2.5 times lower than
those in the remaining three samples.

-- The intrinsic size distributions in all the samples are comparable, so the
differences in the size mean deviations are not statistically significant.

-- The lower limiting angular size and flux density in our sample result in
detections of more distant radio sources (the highest median redshift) than 
those detected in the other samples.

\noindent
Thus, we conclude that a search for giant candidates, among radio sources smaller
than $\sim$4 arc min and with the 1.4-GHz flux density lower than $\sim$100 mJy,
can result in detections of the most distant giant-size sources. In turn, their
studies can shed more light on the problems of the intergalactic medium (IGM)
homogeneity, its cosmological evolution, voids, etc.

\begin{figure*}[h]
\resizebox{170mm}{!}{\includegraphics{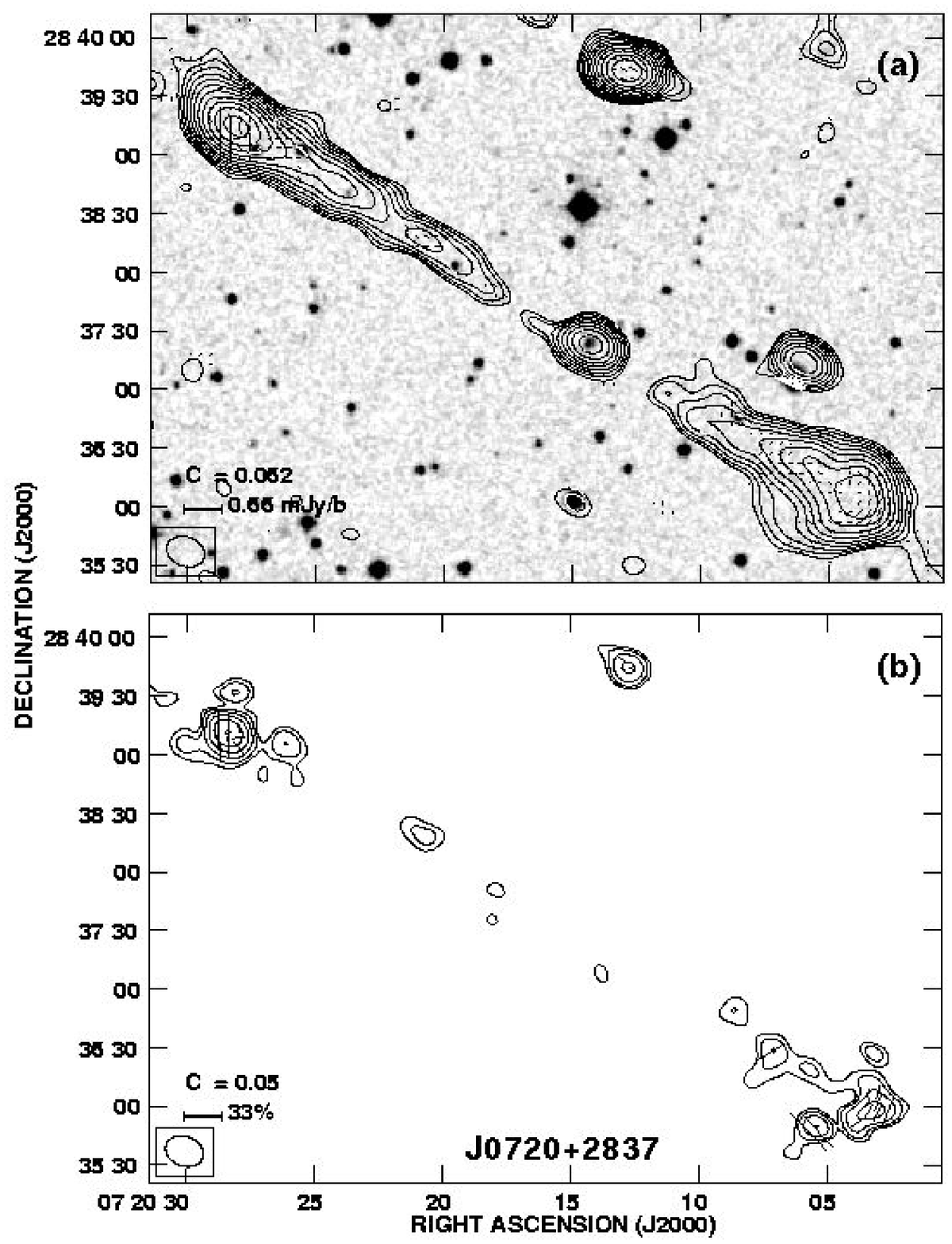}}
\caption{4.86-GHz VLA images of the source J0720+2837. {\bf (a)} total intensity
(Stokes $I$ parameter) logarithmic contours spaced by a factor $2^{1/2}$ are
plotted starting with a value for $C$. The value of $C$ in mJy/beam is given in
each panel and is about 3 times the $rms$ noise level of the particular map.
Superimposed are electric field $E$-vectors (rotated by 90$\degr$) with their
length proportional to the polarised intensity $S_{\rm p}$. The radio contours
are overlaid on the optical field taken from the DSS. Compact
radio core coincidences with the host galaxy. {\bf (b)} Linearly polarised
intensity contours with the vectors of the fractional linear polarisation
superimposed. Lengths of these vectors are proportional to the percent of
polarisation}
\end{figure*}
 
\begin{acknowledgements}
 
The authors acknowledge (i) the National Radio Astronomy Observatory (NRAO)
(Socorro, NM) and Barry Clark for the target-of-opportunity observing time. NRAO
is operated by Associated Universities, Inc., and is a facility of the 
National Science Foundation (NSF), (ii) the Asiago Astrophysical Observatory
(Cima Ekar, Italy) and the Apache Point Observatory  (TX) for the observing time,
and (iii) the Space Telescope Science Institute for the use of the Digitized
Sky Surveys (DSS) data base. This work was supported in part by the State with
funding for scientific research in years 2005-2007 under contract No.
0425/PO3/2005/29.  
 
\end{acknowledgements}

\begin{figure*}[t]
\resizebox{200mm}{!}{\includegraphics{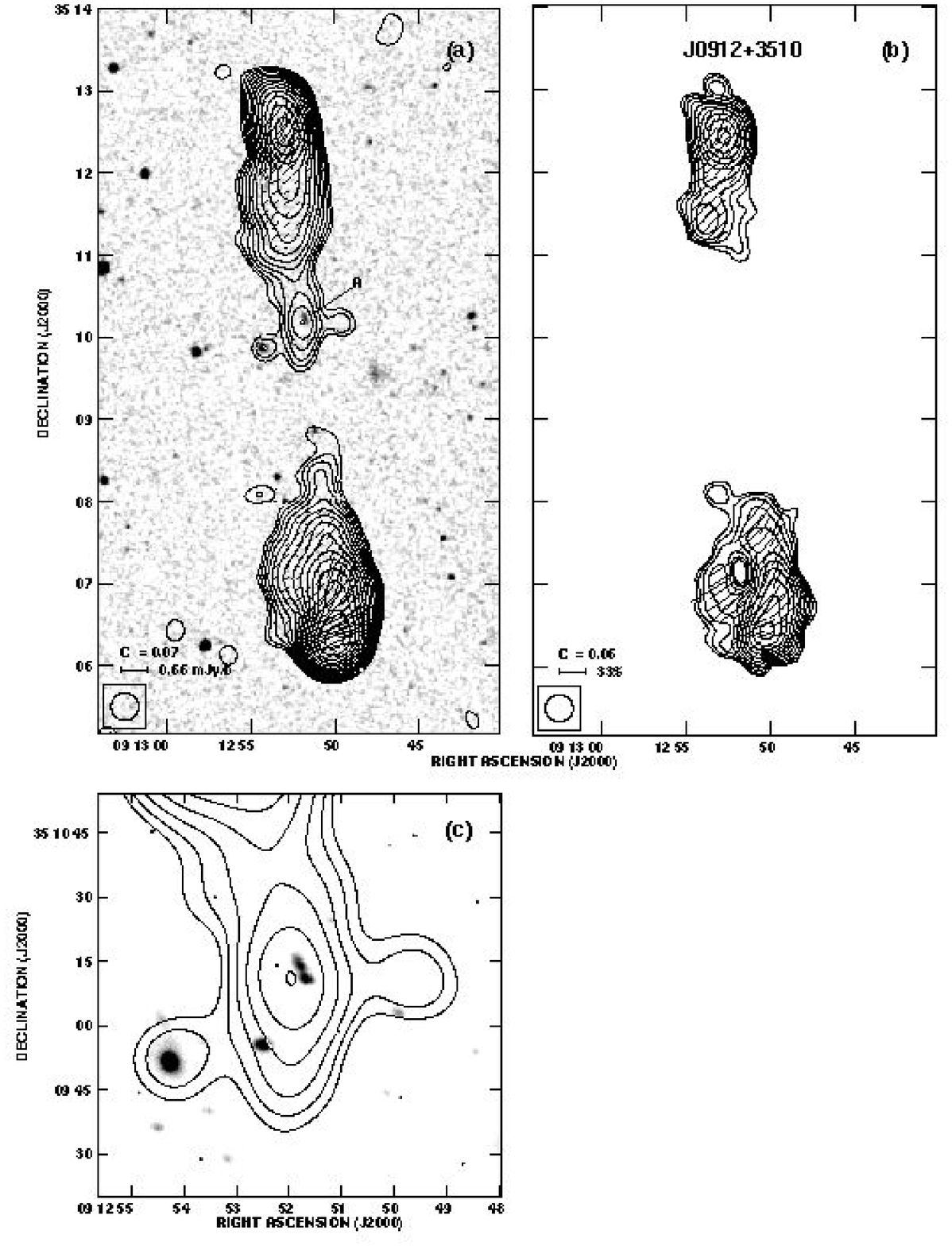}}
\caption{As in Fig.\,3a,b but for the source J0912+3510. Enlarged deep optical
field, taken with the McDonald 2.1m telescope and revealing the close pair of
galaxies coincident with the radio core, is shown panel {\bf (c)}. The full
square in panel {\bf (b)} indicates the position of the evident hot spot}
\end{figure*}

\begin{figure*}[h]
\resizebox{160mm}{!}{\includegraphics{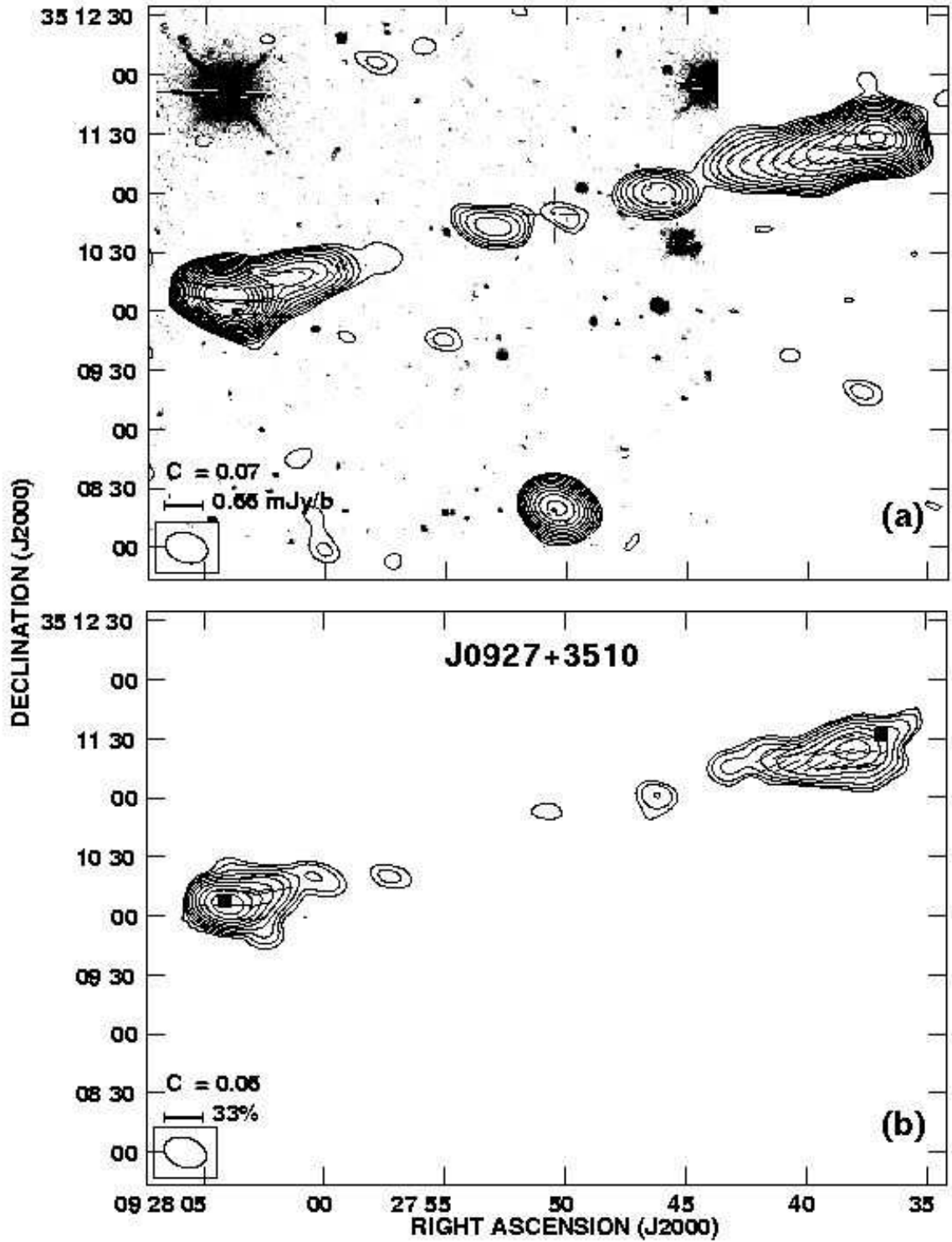}}
\caption{As in Fig.\,3a,b but for the source J0927+3510. The total-intensity
contours in panel {\bf (a)} are overlaid onto the deep optical field taken
with the McDonald 2.1m telescope. The cross marks position of the identified
host galaxy. The full squares in panel {\bf (b)} on this and other maps
indicate the positions of the evident hot spots}
\end{figure*}

\begin{figure*}[t]
\resizebox{130mm}{!}{\includegraphics{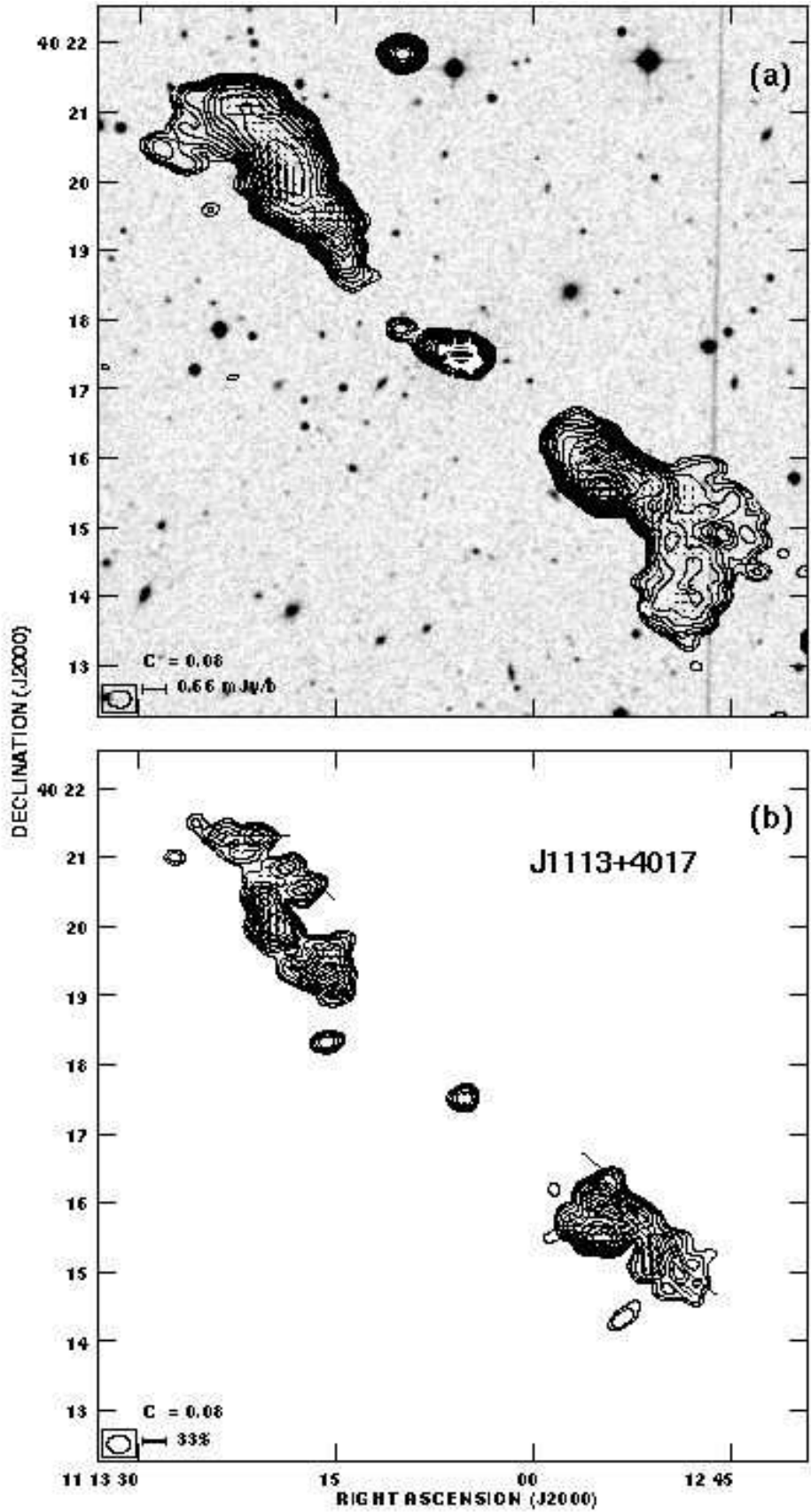}}
\hfill
\parbox{45mm}{
\caption{As in Fig.\,3a,b but for the source J1113+4017}}
\end{figure*}

\begin{figure*}[t]
\resizebox{190mm}{!}{\includegraphics{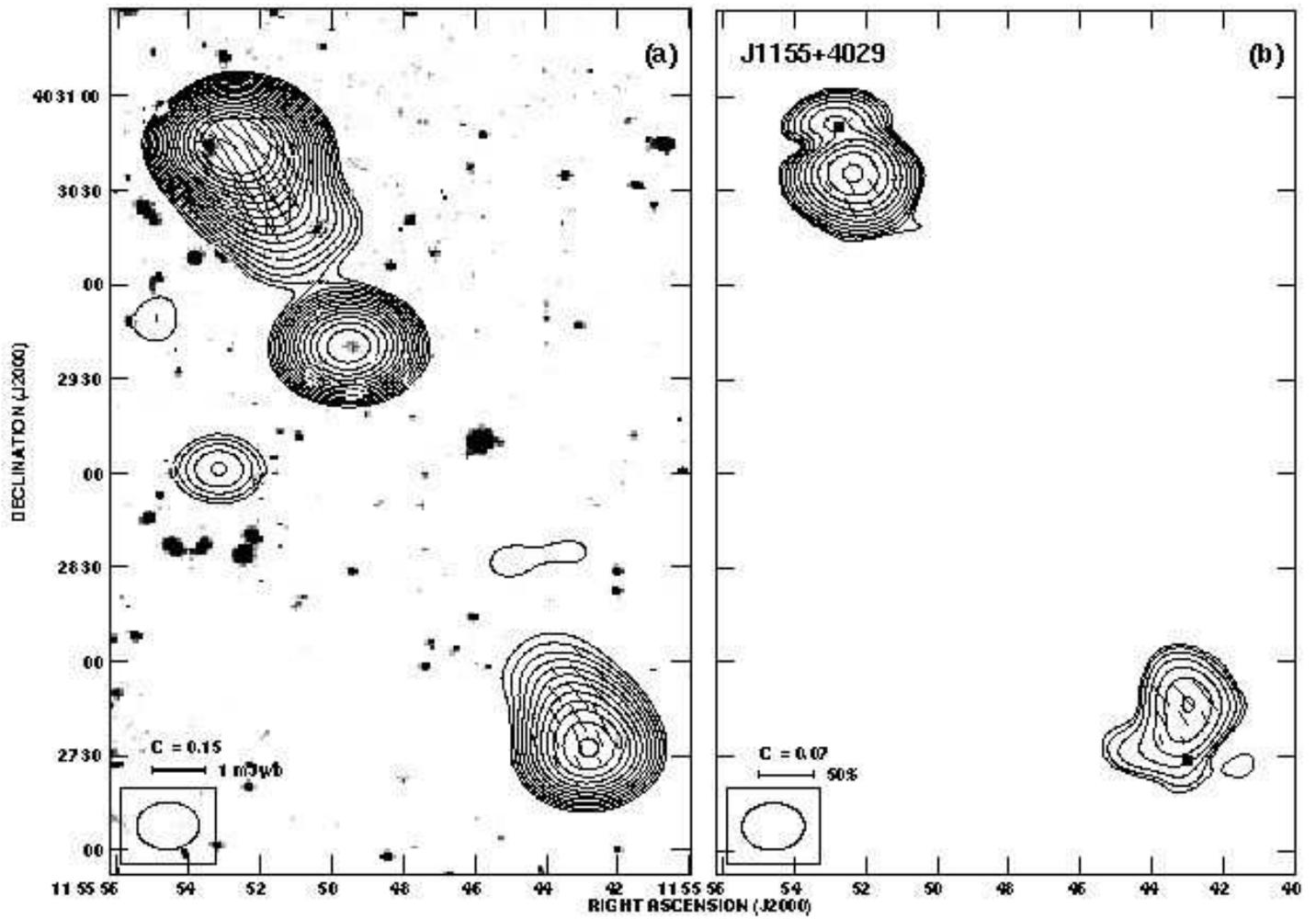}}
\caption{As in Fig.\,3a,b but for the source J1155+4029. The
total-intensity contours in {\bf (a)} are overlaid onto the
deep optical field taken with the McDonald 2.1m telescope
and showing the $R\approx$21.5 mag host galaxy}
\end{figure*}

\begin{figure*}[t]
\resizebox{200mm}{!}{\includegraphics{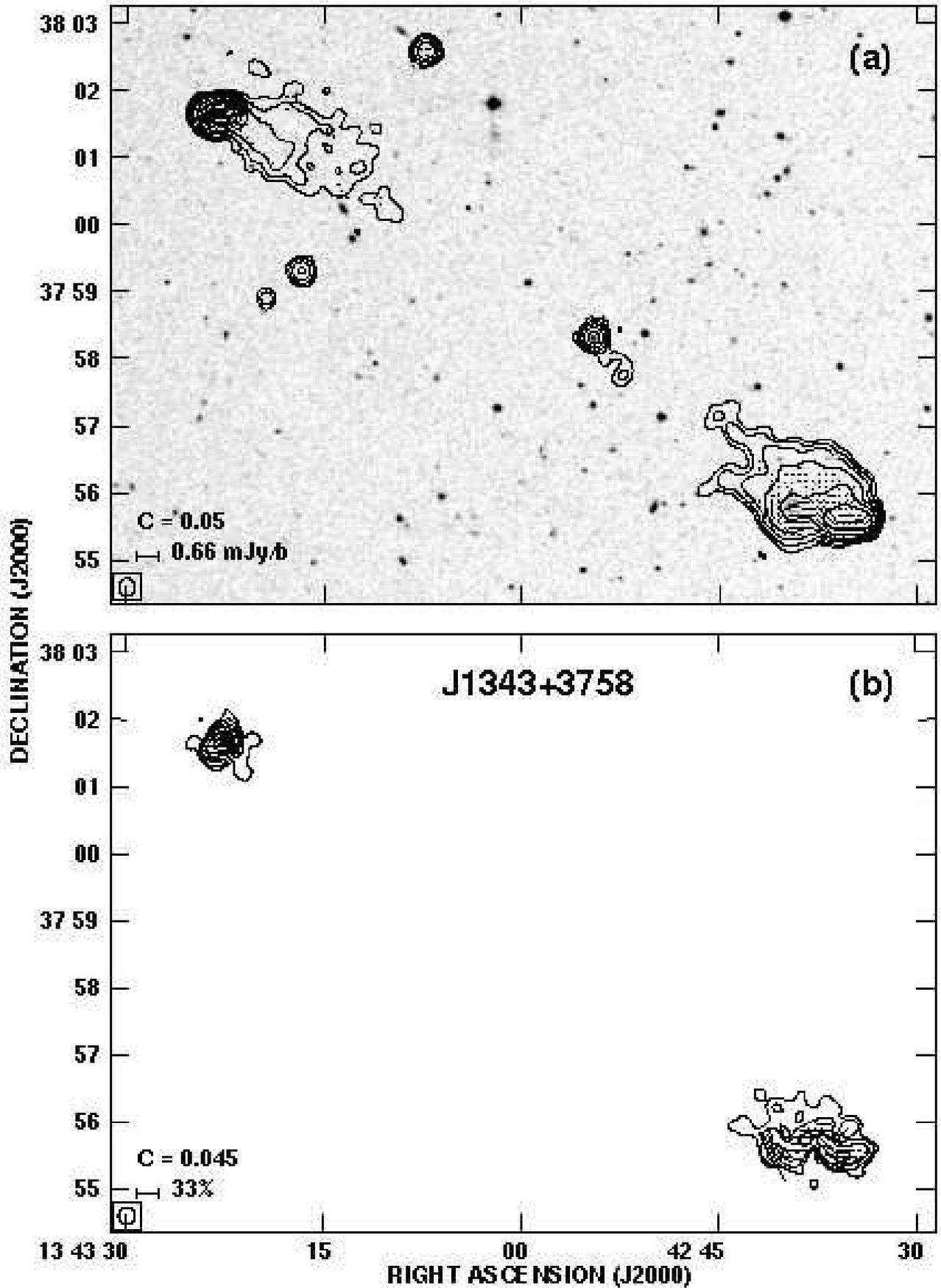}}
\caption{As in Fig.\,3a,b but for the source J1343+3758 except the
logarithmic contours, which are spaced by a factor of 2}
\end{figure*}

\begin{figure*}[t]
\resizebox{200mm}{!}{\includegraphics{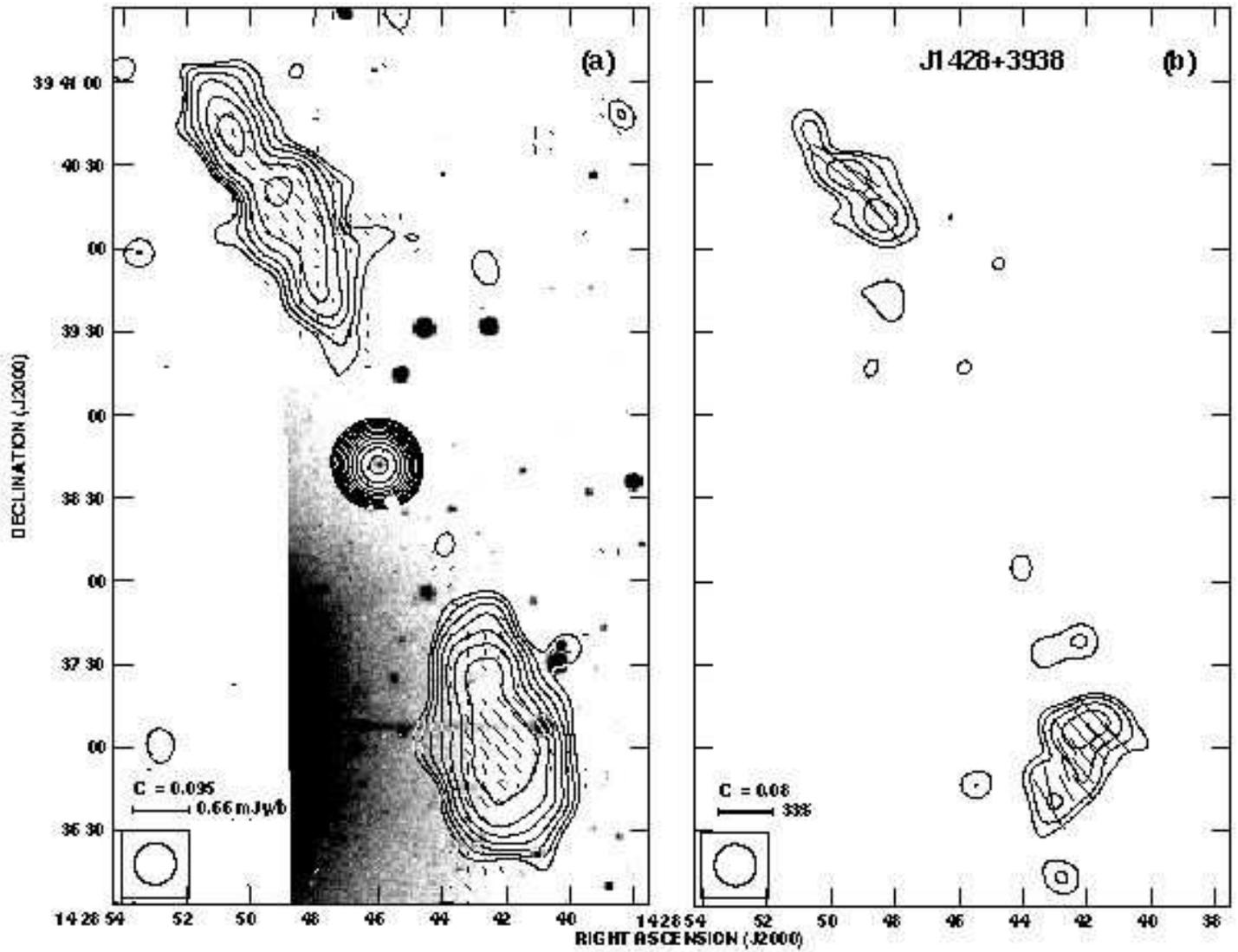}}
\caption{As in Fig.\,3a,b but for the source J1428+3938. The
total-intensity contours in {\bf (a)} are overlaid onto the
deep optical field taken with the Asiago 1.8m telescope
and showing the $R$=21.11 mag host galaxy in the vicinity of a
very bright foreground star}
\end{figure*}

\begin{figure*}[t]
\resizebox{190mm}{!}{\includegraphics{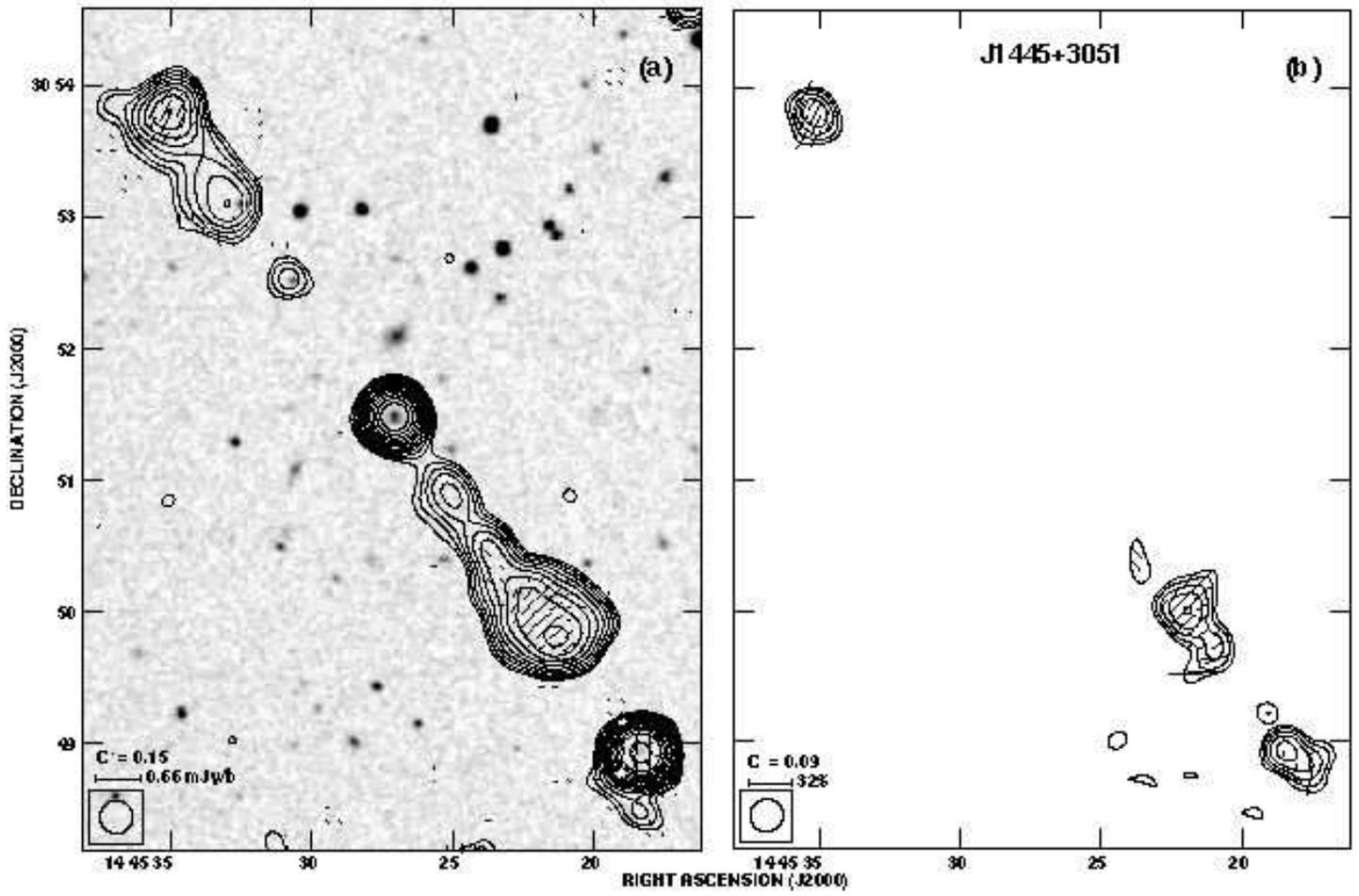}}
\caption{As in Fig.\,3a,b but for the source J1445+3051}
\end{figure*}

\begin{figure*}[t]
\resizebox{130mm}{!}{\includegraphics{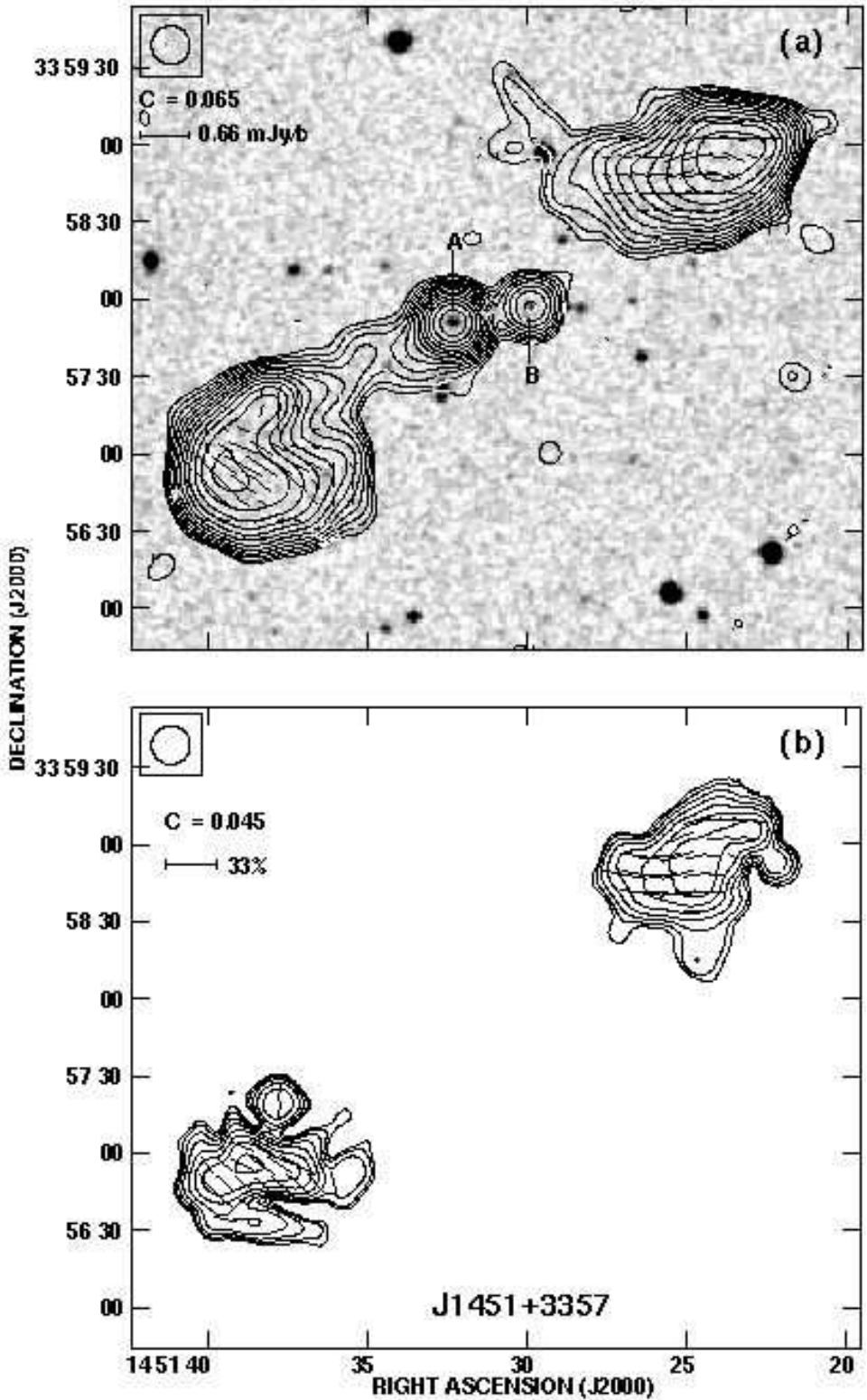}}
\hfill
\parbox{45mm}{
\caption{As in Fig.\,3a,b but for the source J1451+3357. The bright
radio core and the host galaxy are marked with `A'. The compact
component (marked with `B') is a background 0.69 mJy source}}
\end{figure*}

\begin{figure*}[t]
\resizebox{190mm}{!}{\includegraphics{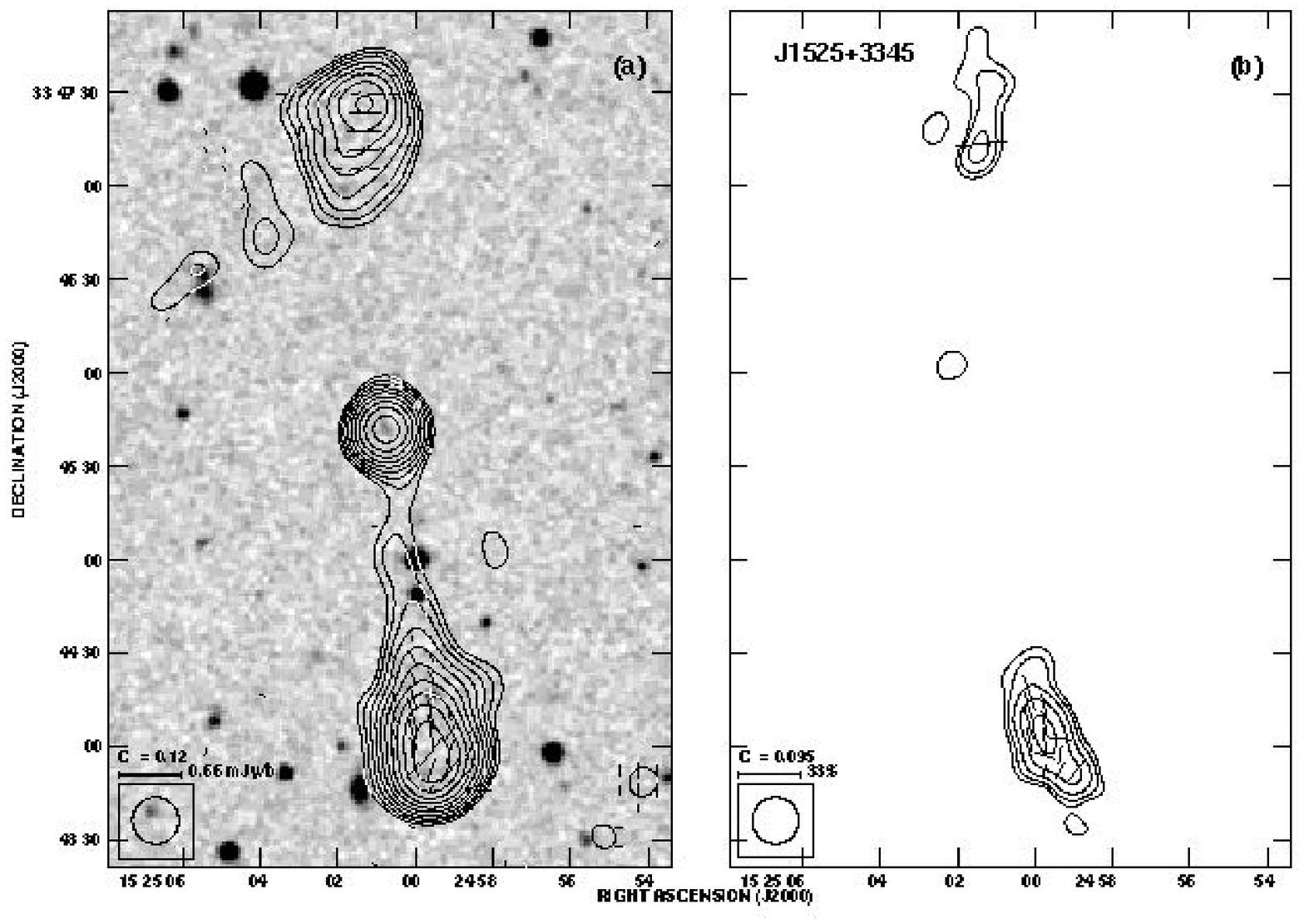}}
\caption{As in Fig.\,3a,b but for the source J1525+3345}
\end{figure*}

\begin{figure*}[thb]
\resizebox{190mm}{!}{\includegraphics{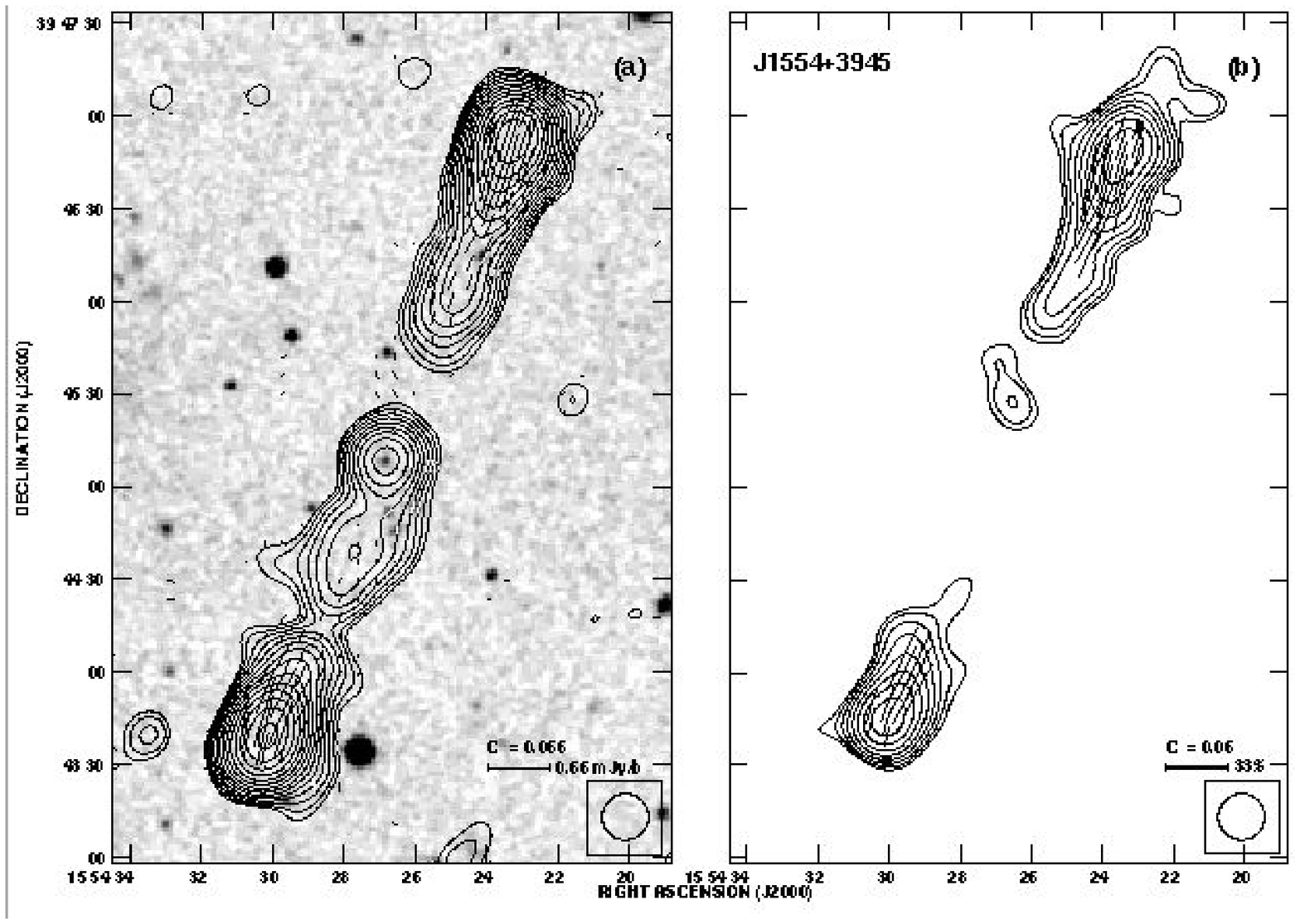}}
\caption{As in Fig.\,3a,b but for the source J1554+3945}
\end{figure*}

\begin{figure*}[thb]
\resizebox{200mm}{!}{\includegraphics{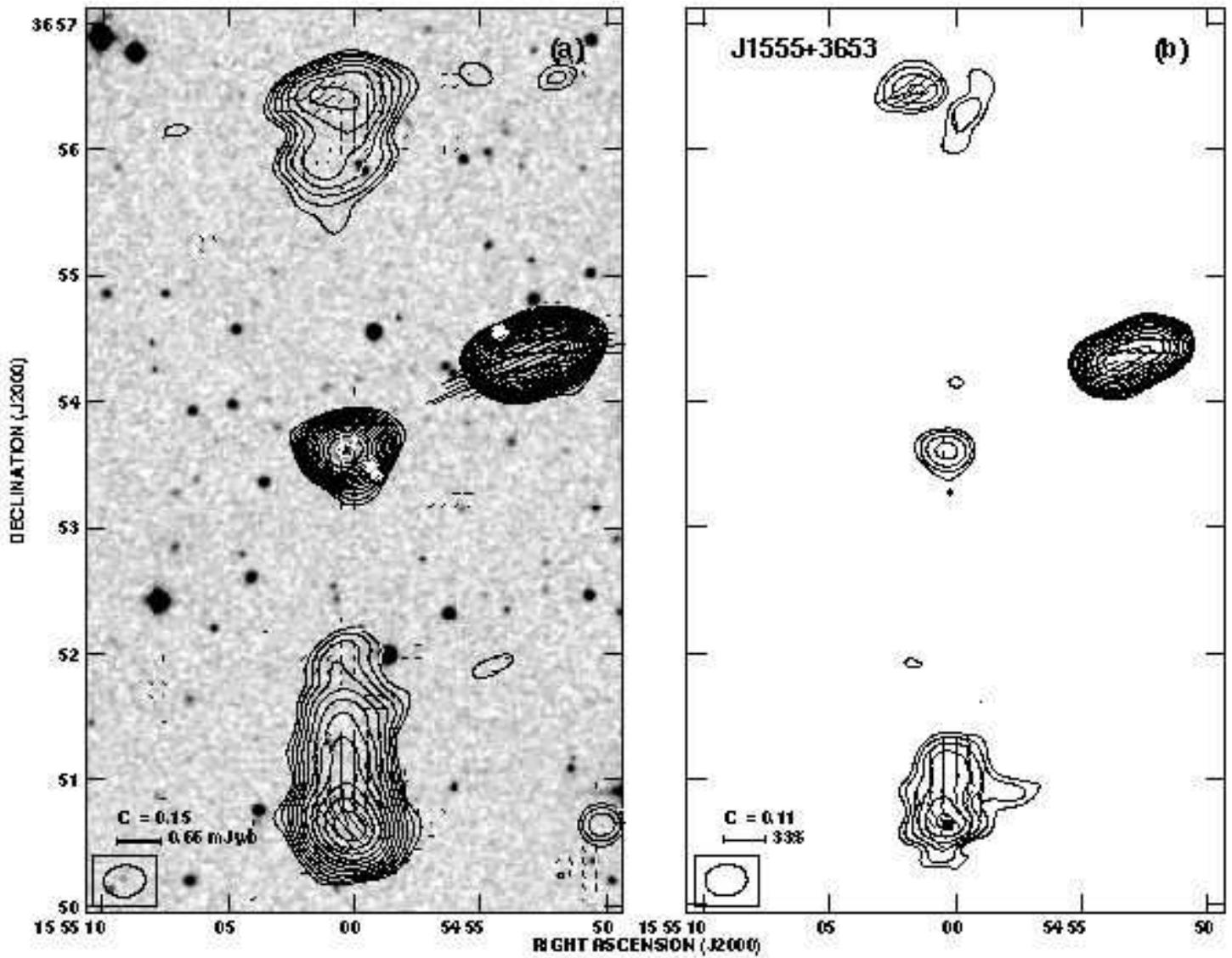}}
\caption{As in Fig.\,3a,b but for the source J1555+3653.
A strong compact double source to the west is unrelated to
the sample source}
\end{figure*}

\begin{figure*}[thb]
\resizebox{190mm}{!}{\includegraphics{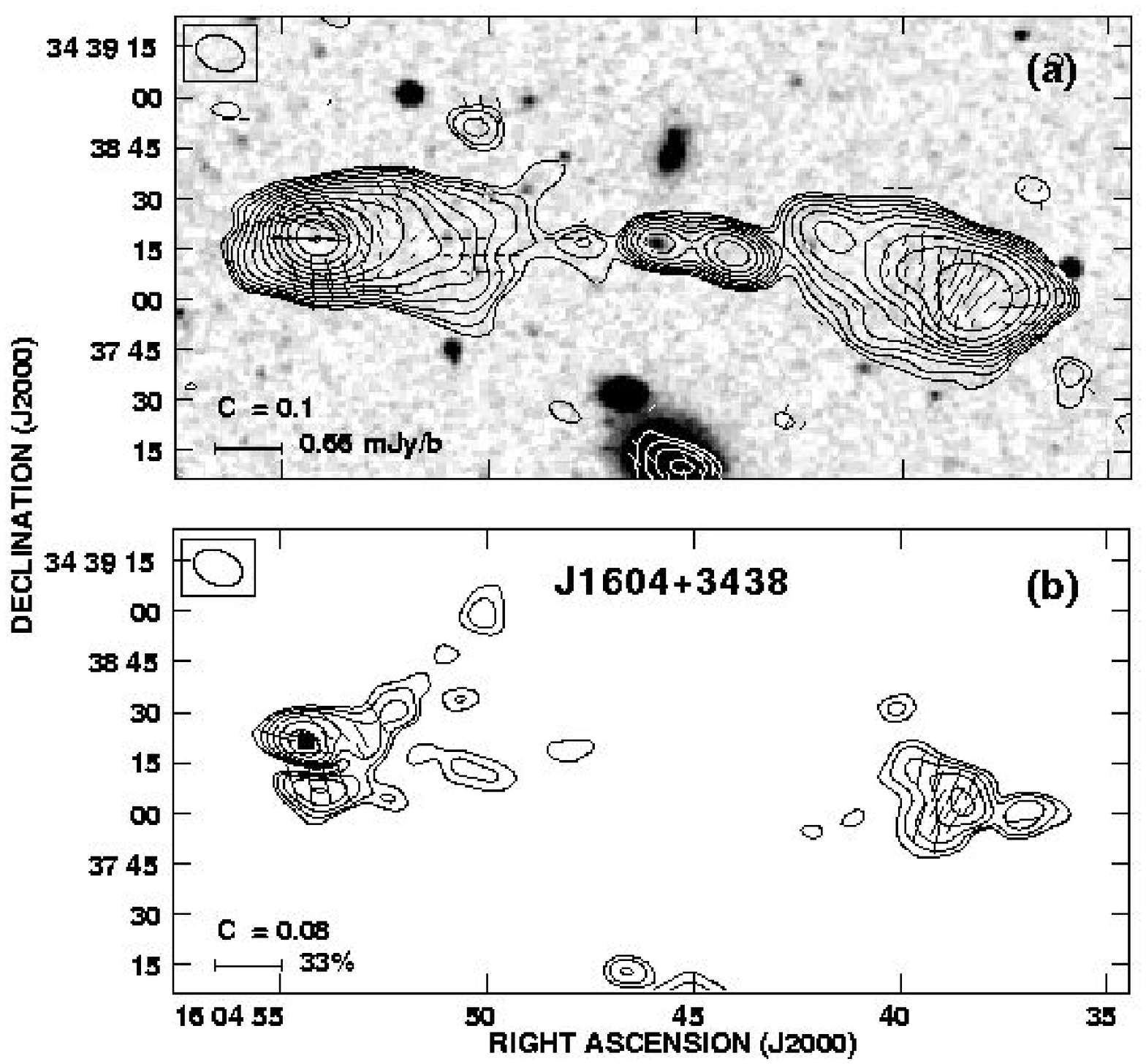}}
\caption{As in Fig.\,3a,b but for the source J1604+3438}
\end{figure*}

\begin{figure*}[thb]
\resizebox{190mm}{!}{\includegraphics{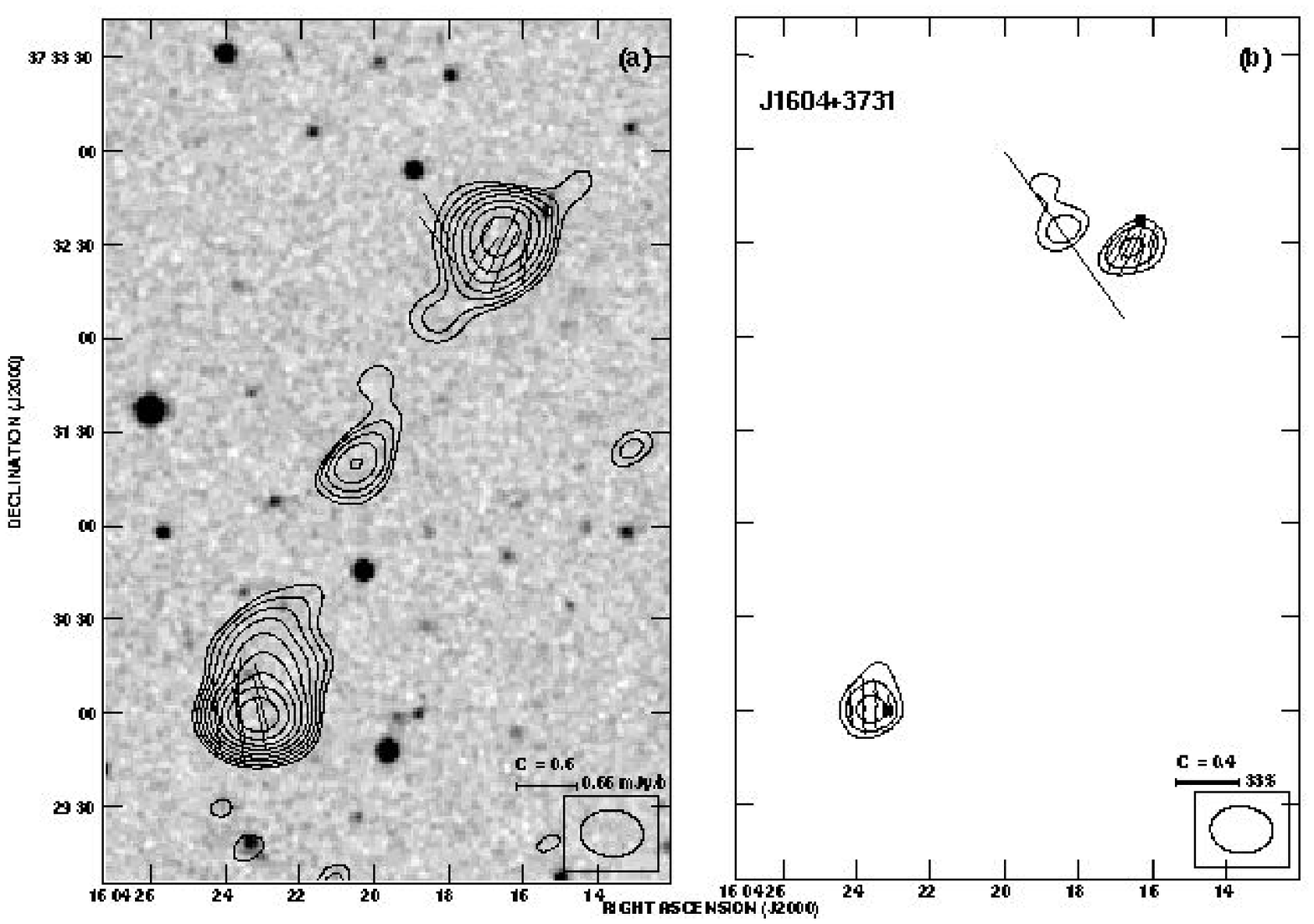}}
\caption{As in Fig.\,3a,b but for the source J1604+3731}
\end{figure*}

\begin{figure*}[thb]
\resizebox{190mm}{!}{\includegraphics{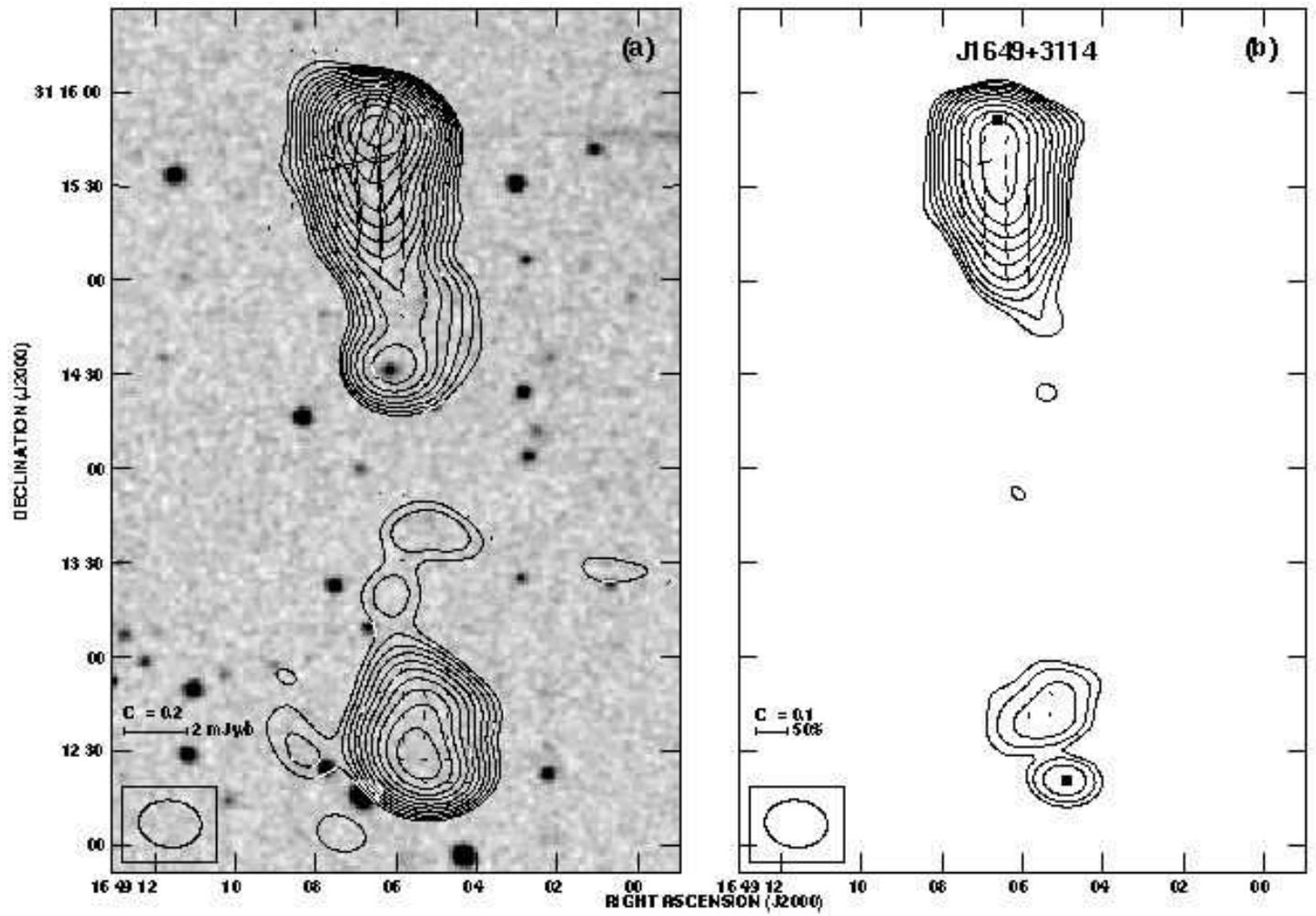}}
\caption{As in Fig.\,3a,b but for the source J1649+3114}
\end{figure*}

\clearpage
\begin{figure*}[thb]
\resizebox{130mm}{!}{\includegraphics{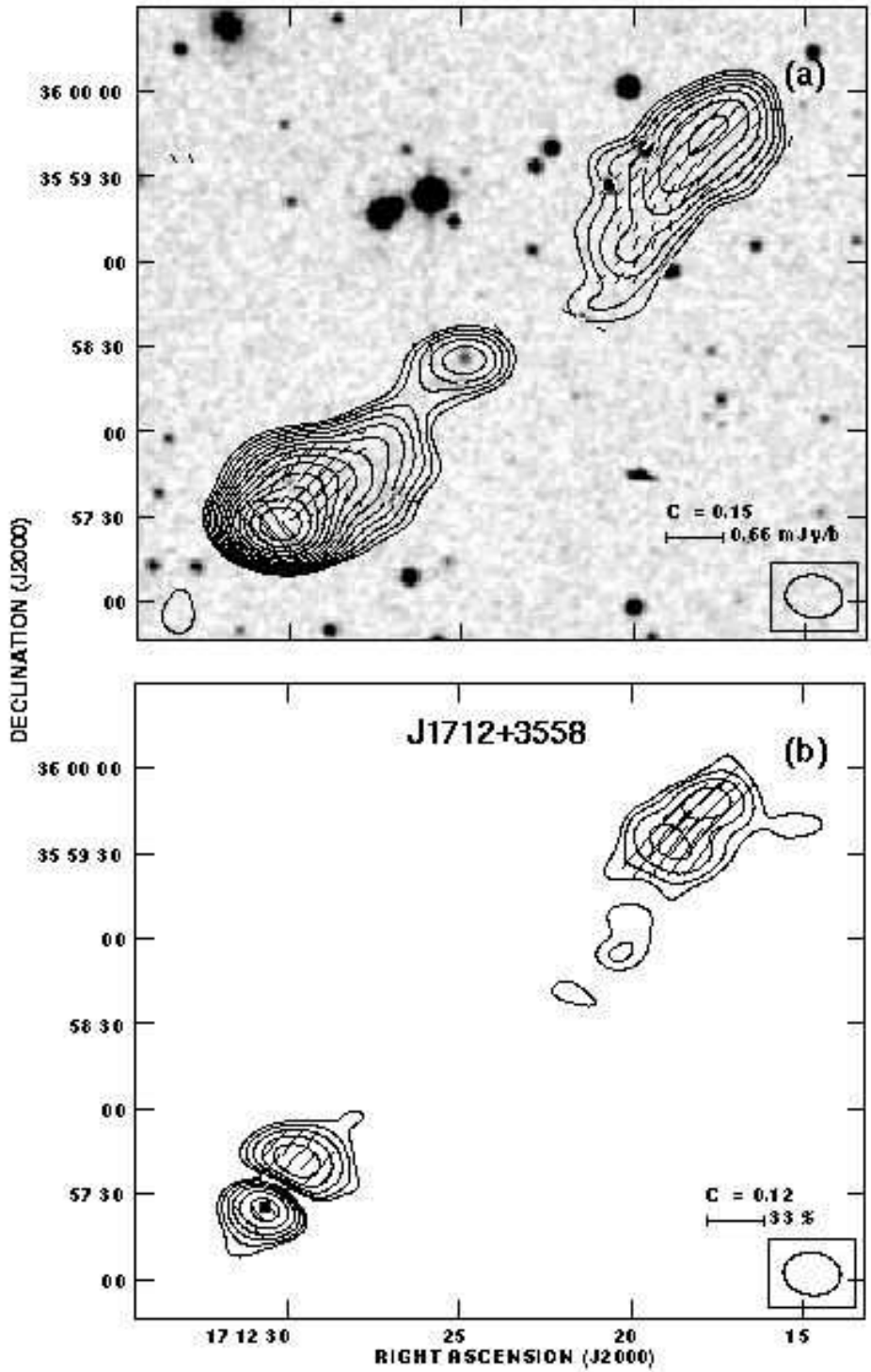}}
\hfill
\parbox{45mm}{
\caption{As in Fig.\,3a,b but for the source J1712+3558}}
\end{figure*}

\begin{figure*}[thb]
\resizebox{190mm}{!}{\includegraphics{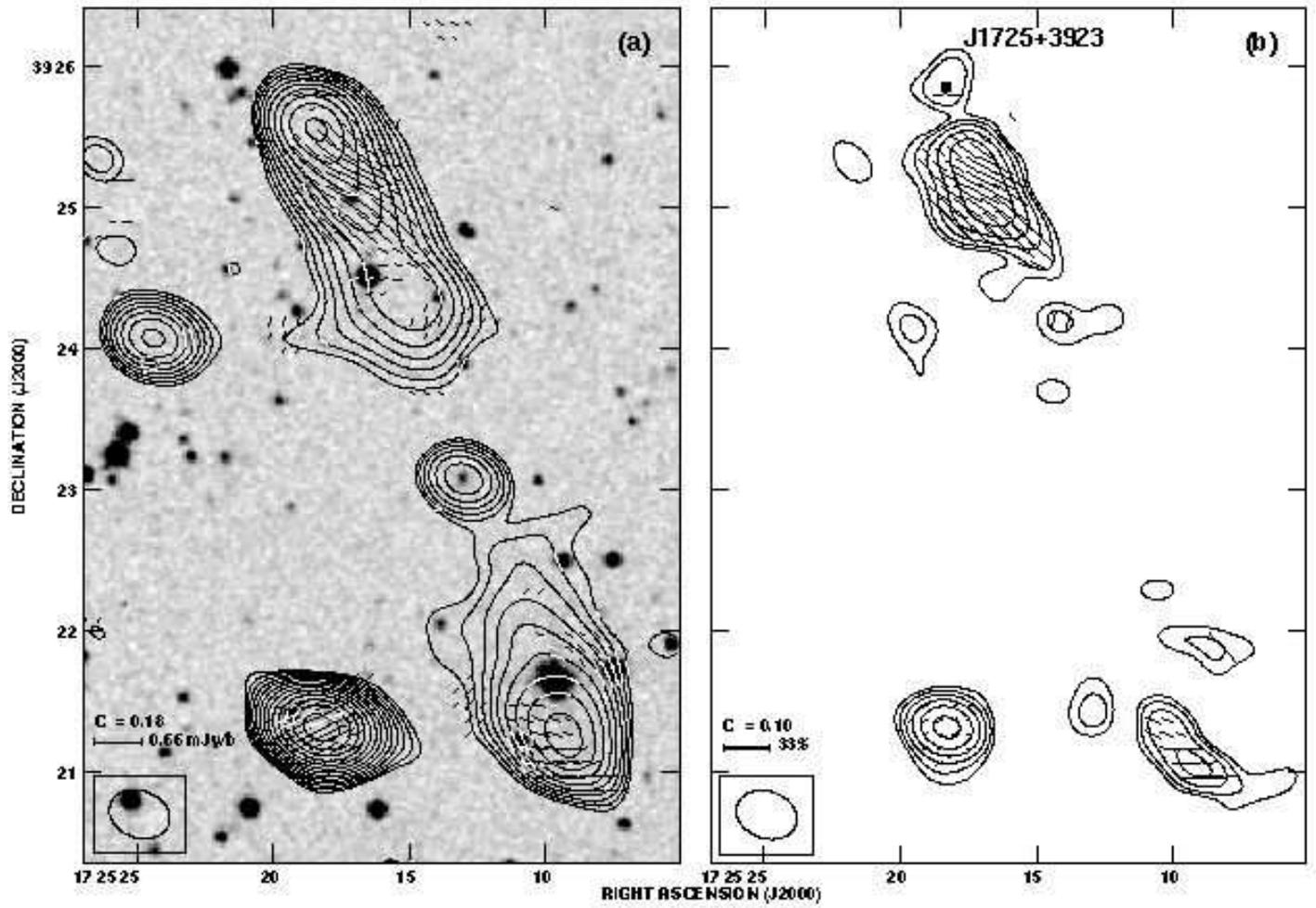}}
\caption{As in Fig.\,3a,b but for the source J1725+3923}
\end{figure*}

\begin{figure*}[thb]
\resizebox{120mm}{!}{\includegraphics{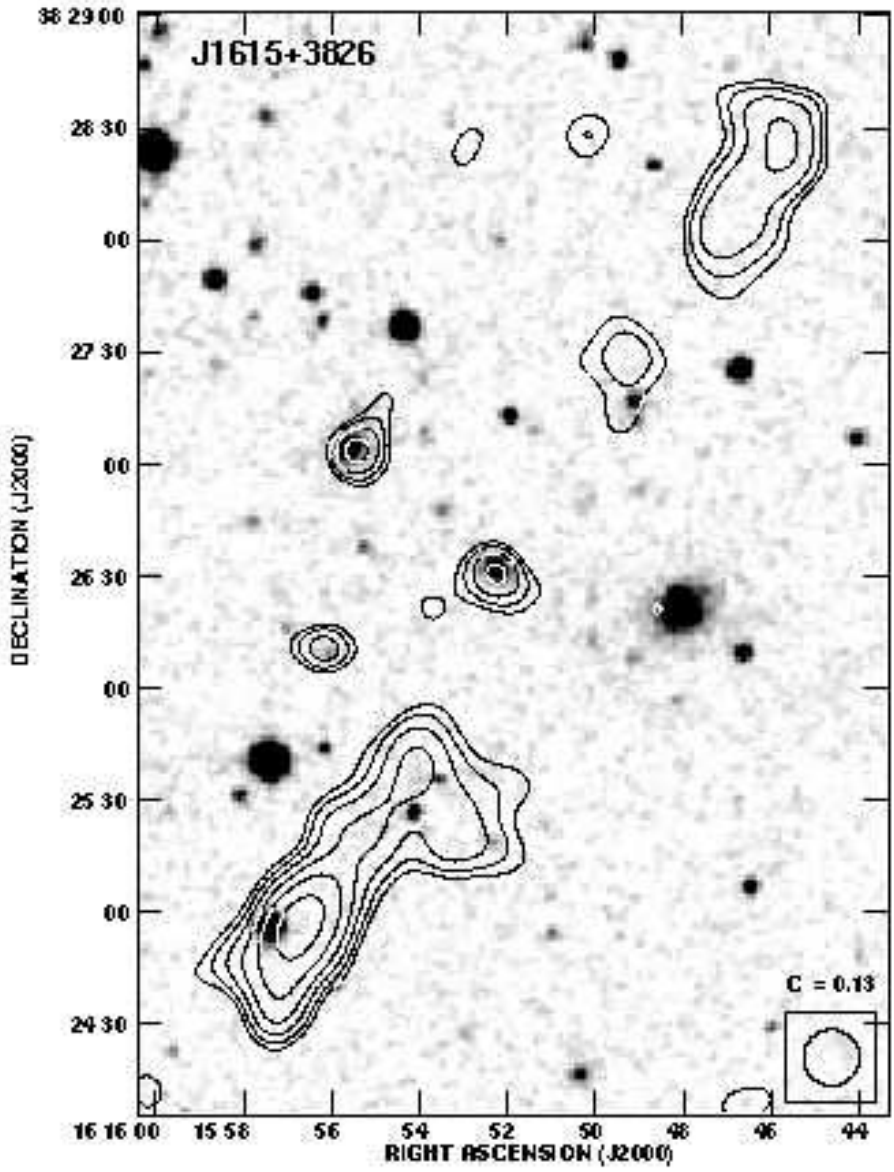}}
\hfill
\parbox{55mm}{
\caption{As in Fig.\,3a but for the source J1615+3826. There is
a lack of detectable polarisation in this source}}
\end{figure*}

\clearpage
\begin{figure}[t]
\resizebox{\hsize}{!}{\includegraphics{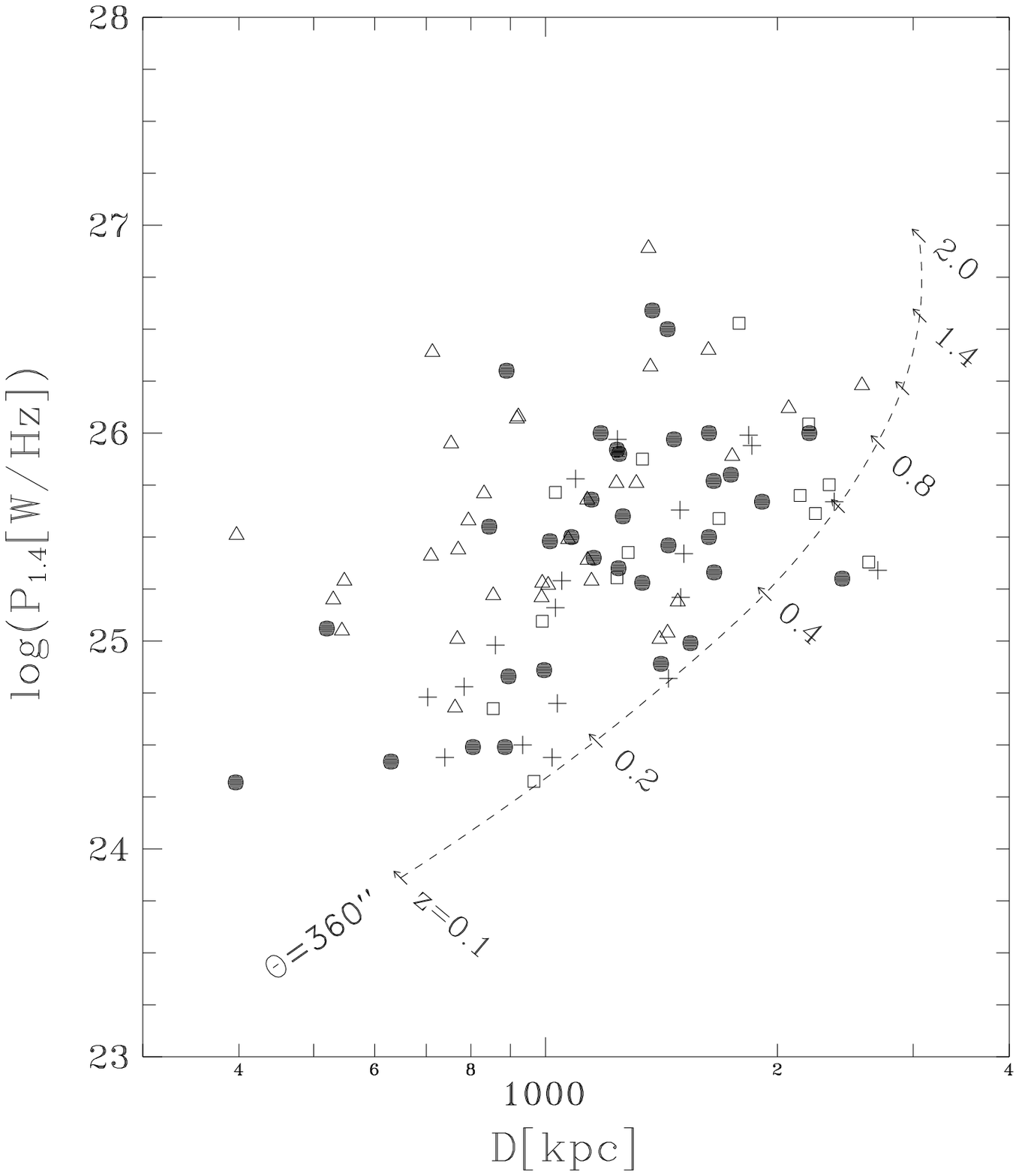}}
\caption{Power vs. size diagram for the samples being compared. Filled circles
mark sources from our sample; triangles correspond to the Lara et al.'s sample;
crosses  to the Schoenmakers et al.'s sample, and open squares  to the Saripalli
et al.'s sample. The dashed curved line indicates the position of a source with the
flux density of 30 mJy and an angular size of 6$\arcmin$ at different redshifts
(marked with small arrows).}
\end{figure} 


\begin{thebibliography}{}
 
\bibitem[1987]{becker}
Becker, R.H., White, R.L., Helfand, D.J., 1995, ApJ, 450, 559 (FIRST)

\bibitem[1999]{blundell}
Blundell, K.M., Rawlings, S., Willott, C.J., 1999, AJ, 117, 766

\bibitem[1999]{bock}
Bock, D.C.J., Large, M.I., Sadler, E.M., 1999, AJ, 117, 1578

\bibitem[1985]{christian}
Christian, C.A., Adams, M., Butcher, H., et al., 1985, PASP, 97, 363

\bibitem[2005]{chyzy}
Chy\.{z}y, K.T., Jamrozy, M., Kleinman, S.J., et al., 2005, Baltic Astr, 14, 358

\bibitem[1998]{condon}
Condon, J.J., Cotton, W.D., Greisen, E.W., et al., 1998, AJ, 115, 1693 (NVSS)

\bibitem[1996]{cotter}
Cotter, G., Rawlings, S., Saunders, R., 1996, MNRAS, 281, 1081

\bibitem[1976]{cousins}
Cousins, A.W.J., 1976, MNRAS, 81,25

\bibitem[1989]{fr}
Fanaroff, B.L. \& Riley, J.M., 1974, MNRAS, 167, 31P

\bibitem[1985]{ficarra}
Ficarra, A., Gruef, G., Tomassetti, G., 1985, A\&AS, 59, 255

\bibitem[2001]{giovan}
Giovannini, G., Cotton, W.D., Feretti, L., et al., 2001, ApJ, 552, 508

\bibitem[1998]{hook}
Hook, I.M., Becker, R.H., McMahon, R.G., White, R.L., 1998, MNRAS, 297, 1115

\bibitem[2005]{jamrozy}
Jamrozy, M., Machalski, J., Mack, K.-H., Klein, U., 2005, A\&A, 433, 467

\bibitem[1997]{kda}
Kaiser, C.R., Dennett-Thorpe, J. \& Alexander, P., 1997, MNRAS, 292, 723

\bibitem[1992]{kenn}
Kennicutt, R.C., 1992, ApJS, 79, 255

\bibitem[2004]{konar}
Konar, C., Saikia, D.J., Ishwara-Chandra, C.H., Kulkarni, V.K., 2004, MNRAS, 355, 845

\bibitem[2004]{kron}
Kronberg, P.P., Colgate, S.A., Li, H., Dufton, Q.W., 2004, ApJ, 604, L77

\bibitem[1984]{laing}
Laing, R.A., 1984, in `Physics of Energy Transport in Extragalactic Radio Sources',
Green Bank (NRAO), eds.: Bridle, A.H. \& Eilek, J.A., p. 90

\bibitem[2001]{lara}
Lara, L., Cotton, W.D., Feretti, L., et al., 2001, A\&A, 370, 409

\bibitem[1991]{leahy}
Leahy, J.P. \& Williams, A.G., 1984, MNRAS, 210, 929

\bibitem[2001]{mjz}
Machalski, J., Jamrozy, M., Zola, S., 2001, A\&A, 371, 445 (Paper I)

\bibitem[2002]{manolakou}
Manolakou, K. \& Kirk, J.G., 2002, A\&A, 391, 127

\bibitem[1980]{miley}
Miley, G.K., 1980, ARA\&A, 18, 185

\bibitem[1997]{rengelink}
Rengelink, R.B., Tang, Y., de Bruyn, A.G., et al., 1997, A\&AS, 124, 259 (WENSS)

\bibitem[2005]{saripalli}
Saripalli, L., Hunstead, R.W., Subrahmanyan, R., Boyce, E., 2005, AJ, 130, 896

\bibitem[2001]{schoen}
Schoenmakers, A.P., de Bruyn, A.G., R\"{o}ttgering, H.J.A., van der Laan, H.,
   2001, A\&A, 374, 861
\end{thebibliography}
\end{document}